\documentclass[reprint,amsmath,amssymb,aps]{revtex4-1}

\usepackage{graphicx}
\usepackage{bm}
\usepackage{physics}
\usepackage{natbib}
\usepackage{amsfonts,amsthm,bbold}
\usepackage{tikz}
\usepackage{mathtools}
\usepackage[justification=justified]{caption}

\begin{document}

\preprint{APS/123-QED}

\title{Burning the Trojan Horse: \\ Defending against Side-Channel Attacks in QKD}

\author{Scott E. Vinay}
 \email{svinay1@sheffield.ac.uk}
\author{Pieter Kok}
% \email{p.kok@sheffield.ac.uk}
\affiliation{ Department of Physics and Astronomy, University of Sheffield}

\date{\today}

\begin{abstract}
\noindent The discrete-variable QKD protocols based on BB84 are known to be secure against an eavesdropper, Eve, intercepting the flying qubits and performing any quantum operation on them. However, these protocols may still be vulnerable to side-channel attacks. We investigate the Trojan-Horse side-channel attack where Eve sends her own state into Alice's apparatus and measures the reflected state to estimate the key. We prove that the separable coherent state is optimal for Eve amongst the class of multi-mode Gaussian attack states. We describe how Alice may defend against this by adding thermal noise to the system and give an analytic expression of the resulting secret key rate. We also provide a bound on the secret key rate in the case where Eve may use any separable state, and describe an active defense system based on optical modulators.
\end{abstract}

\maketitle

\section{Introduction}

\noindent Quantum Key Distribution (QKD) systems are generally created with the promise that the uncertainty inherent in quantum measurements allows for two or more parties to communicate with unconditional security. By this, it is meant that an eavesdropper, Eve, may be imbued with unbounded computational power and be able to do anything that is allowed by the laws of physics, yet still only achieve a level of mutual information with a bit string shared by valid parties Alice and Bob that is exponentially small with the key length \cite{Scarani2009TheDistribution}. This is in contrast with classical encryption, for which the above claim only holds when Eve has some bounded computational ability (which may be exceeded by a quantum computer \cite{Shor1997Polynomial-TimeComputer}).

In any claim of security assumptions will necessarily be made on restrictions on the methods by which Eve may try to learn the key. For example, it is clear that if Eve has unrestricted access to Alice's lab then no level of sophistication in the protocol can prevent her from learning the key. Therefore we must always decide on a boundary demarcating the quantum or classical objects that Eve may access from the ones that she may not.

In the standard proofs of the security of many QKD protocols, it is assumed that Eve may interact with any of the ``flying'' photonic qubits that are sent between Alice and Bob and with the quantum channel carrying them. Her operations may include storing the qubits for arbitrarily long periods of time, performing multipartite rotations or measurements, entangling these with ancilla states, or replacing sections of the channel with loss-free channels. Renner used the de Finetti theorem to show that this very general case is equivalent to the case of Eve performing operations on one qubit at a time \cite{renner_replaced}, and along with \cite{biham1997bounds} and \cite{Inamori2007UnconditionalDistribution}, this proves the security of BB84 \cite{Bennett1984QuantumTossing.} against such attacks.

However, we must assume that Eve is wily and cunning, and will seek alternative hidden avenues known as \emph{side-channel attacks} (SCAs) \cite{Newton2015NovelDistribution,lydersen2010hacking,Comandar2016QuantumLasers,scarani2004quantum_replace,Curty2014Finite-keyDistribution,Braunstein2012Side-channel-freeDistribution,Gerhardt2011Full-fieldSystem,Xu2010ExperimentalSystem,Zhao2008QuantumSystems,Wang2005BeatingCryptography}. One such SCA that has recently received theoretical \cite{gisin2006trojan,deng2005improving,Lucamarini2015PracticalDistribution} and experimental \cite{jain2015risk,jain2014trojan,Sajeed2015AttacksTossing} attention is the so-called \emph{Trojan Horse Attack} (THA). Here, Eve will tap into the optical channel that Alice and Bob use to communicate. She will then send her own optical state into Alice's system, whereupon it will reflect off the same apparatus used to encode the legitimate photonic qubits. Having picked up some information on the encoding of the latest quantum state that Alice sent, it will return out and be measured by Eve. Eve will then use the result of this measurement, possibly combined with some operation on the legitimate qubits, to make a best estimate of the state that Alice sent to Bob, thus giving her some non-negligible information mutual with the key.

This attack has previously been analysed by Lucamarini et al. \cite{Lucamarini2015PracticalDistribution} They assume that Eve uses a coherent state to probe the system, and describe using a one-way attenuating filter at the entry-point of Alice's apparatus as a defense. The effect of this is to absorb the majority of light that is sent into the system, such that Eve receives far less than one photon back per attempt, reducing her ability to estimate the key bit. They make use of the theoretical framework of Gottesman et al. \cite{gottesman_replaced} to get an expression for the rate at which Alice and Bob can generate a secret key in the presence of such an attack.

In this paper we make use of this same framework, but extend the range of powers of both Eve and Alice. The paper is organised as follows. In section \ref{sec:prelim} we describe some of the fundamental notions necessary to understand the process of encoding in a phase-modulated BB84 protocol. In section \ref{sec:thermalnoise}, we describe and analyse the effect of Eve performing a THA on the system, allowing her to use any Gaussian state including multimode entangled states. We prove that the (separable) coherent states are optimal amongst this class. As part of this analysis, we describe a defense system that Alice may employ that involves deliberately adding thermal noise to the system. We show that this can more than double the range over which a QKD system can securely communicate. 

Motivated by the revelation that entanglement does not assist Eve when using Gaussian states to attack the system, in section \ref{sec:separable} we restrict Eve to separable states. We derive a bound on the information that Eve may learn about the key when we allow her to use \emph{any} separable state. Finally, in section \ref{sec:shutter} we describe an active defense system that may be used in place of attenuating filter. This relies on the causing Eve's THA light pulse to undergo many reflections before being returned.

\section{Preliminary notions} \label{sec:prelim}

\begin{figure}[t]
   \centering
     \includegraphics[width=0.8\columnwidth]{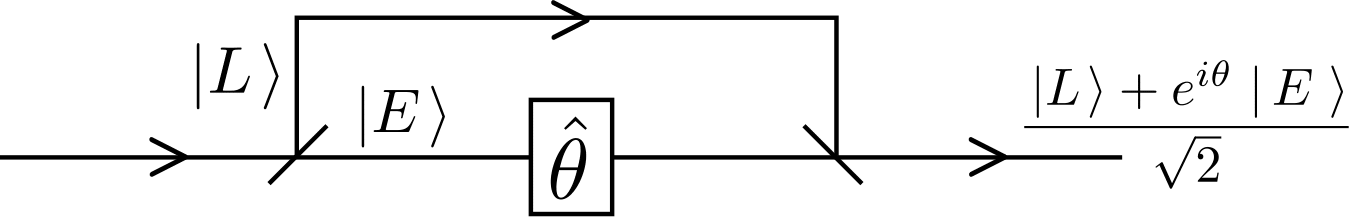}
     \captionsetup{justification=justified}
     \caption{A phase-shift between an early and a late mode, parameterised by $\theta$, encodes the quantum bit.}
     \label{fig:earlylate}
\end{figure}

We consider here an implementation of the BB84 \cite{Bennett1984QuantumTossing.} protocol. Here, Alice chooses one of two mutually unbiased bases, $X$ and $Y$. After choosing a basis she then sends a photon encoding either state $\ket{0_{X,Y}}$ or $\ket{1_{X,Y}}$. These states are encoded as

\begin{eqnarray}
\ket{0,1_{X,Y}} = \frac{ \ket{L} + e^{i \theta} \ket{E}}{2},
\end{eqnarray}

\noindent where $\ket{E}$ and $\ket{L}$ are early and late modes as shown in Fig. \ref{fig:earlylate}.

The key parameter here is $\theta$, which encodes the state as follows:

\begin{eqnarray}
\begin{split}
\ket{0_X}& \rightarrow \theta = 0 \hspace{7mm} \ket{0_Y} \rightarrow \theta = \pi/2 \\
\ket{1_X}& \rightarrow \theta = \pi \hspace{6.7mm} \ket{1_Y} \rightarrow \theta = 3\pi/2 
\end{split}
\end{eqnarray}

It is this parameter that Eve wishes to estimate. In order to do this, she prepares her own state, $\rho$. This is assumed to exist in the photonic Fock space of a single mode. The single mode assumption is justified since we may say that Alice will filter out all frequencies that are not equal to the one sent to Bob. It may also be assumed without loss of power or generality to be pure. This state is sent into Alice's system. Here it passes through a filter which allows a fraction $\eta \ll 1$ of the light to be transmitted, resulting in a state $\rho_\eta$. It then reaches the polarizing filter, where it evolves according to the same Hamiltonian that encoded $\theta$ into the photon that was sent to Bob. That is to say, it is transformed as follows:

\begin{equation}
\rho_\eta \rightarrow \rho_\eta^\theta \equiv e^{i \theta \hat{a}^\dag\hat{a}} \rho_\eta \hspace{1mm} e^{-i \theta \hat{a}^\dag\hat{a}},
\end{equation}

\noindent where $\hat{a}$ is the annihilation operator on the Fock space of Eve's photons. After Eve's state has picked up the phase information it returns to her. She then performs some operation to try to make an estimate of $\theta$.

\section{Gaussian state attack and thermal defense} \label{sec:thermalnoise}

Attenuation-based defense systems, which aim to muddy the phase information on Eve's state by blocking most of the incoming attack state, have been previously analysed by Lucamarini et al. \cite{Lucamarini2015PracticalDistribution} They show that in order to realise any appreciable level of security, a very high level of attenuation is required. Specifically it requires that Eve get back far less than one photon per attempt. We want to investigate whether it is possible to relax this requirement by implementing complementary security measures.

To this end, we will consider a system where Alice adds a small amount of thermal noise with average photon number $\mu_T$ into the signal she sends out to Bob. Since Eve taps into this same channel, some of this noise will also be picked up by Eve. It will be combined with her returned state $\rho_\eta^\theta$ to produce the state $\rho_{\eta,\mu_T}^\theta$.

When running a protocol such as BB84, Alice and Bob can only try to generate secret key bits from the attempts where Bob successfully received a signal. So after post-selecting on these bits Bob must receive at least one photon per bit generation attempt. On the other hand, we may choose the strength of the attenuator such that Eve receives much less than one photon per attempt. Therefore, if $\mu_T$ is less than one, but comparable to the average photon number of $\rho_\eta$, then the addition of this thermal noise is likely to affect Eve more than it affects Bob. Subsections \ref{subsec:secretkeyrate} and \ref{subsec:effectOnBob} aim to quantify this.

\subsection{State description} \label{subsec:stateDescription}

Here we will describe specifically how we construct $\rho_{\eta,\mu_T}^\theta$ from $\rho$, and how Eve should choose $\rho$ to maximise her knowledge of the key.

Firstly, it is clear that the choice of initial state $\rho$ will have a significant effect on Eve's ability to discern $\theta$. There are certain properties of this state that we can identify that we expect to affect this in varying degrees. 

The property that may be most apparent is that of the average photon number of the state. If Eve sends in a single photon, then given a high amount of attenuation, she is not likely to get much back and will not be able to reliably learn $\theta$. On the other hand if she is allowed to send in an arbitrarily bright state with unbounded average photon number it is clear that she will always be able to distinguish the different settings of $\theta$ perfectly. Therefore, to be able to implement any QKD protocol, the first step in protecting against a THA is putting some upper bound on the average number of photons that may pass into the system. This may be done by way of some defense such as an optical fuse \cite{Driscoll1991ExplainingFuse.}, which melts when sufficiently many photons pass through it, or by identifying some other component which will be irreversibly damaged when subject to a bright enough light \cite{Carr2003WavelengthMechanisms}. A more detailed examination of the numbers and figures behind such defenses may be found in \cite{Lucamarini2015PracticalDistribution}, but for our purposes we may simply assume that there does exist some bound $N$ such that $\left\langle\smash{\hat{n}}\right\rangle_\rho \equiv \Tr\left[\hat{n}\rho\right] < N$, where $\hat{n} = \hat{a}^\dag\hat{a}$. 

Another relevant property may be the purity of the state $\rho_\eta$ \emph{after} passing through the attenuator. Most states will become mixed after undergoing loss, however coherent states (as used in \cite{Lucamarini2015PracticalDistribution}) will not. They are instead mapped to coherent states with lower photon numbers. As a result of this, the loss does not introduce any \emph{classical} uncertainty into the estimation of the phase. It may also be the case that entanglement assists the estimation, as is the case with entanglement-assisted illumination \cite{Lloyd2008EnhancedIllumination}. It is important that we search for the most powerful possible attack that Eve may make, taking all of these factors into consideration. It is only then that we may have confidence in our security proofs against the THA or other SCAs.

We will consider the case where Eve may use any multi-mode entangled Gaussian state. Since only one mode enters Alice's apparatus, Eve needs only to use at most one idler mode, which she retains as a reference \cite{nielsen2002quantum}. This state is created by applying a two-mode squeezer to the vacuum followed by a displacement on the mode that enters Alice's system (applying a displacement to the idler mode turns out to have no effect on the the amount of information that Eve may learn about the key). Up to a change of variables in the squeezing and displacement parameters, this setup is equivalent to all other combinations of Gaussian operations \cite{Braunstein2005SqueezingResource}, such as applying single-mode squeezers and displacing before squeezing. This, therefore, represents the most general Gaussian-state attack that Eve may make.

Eve's initial state is then 

\begin{equation} \label{eq:defOfState}
\rho = \hat{D}(\alpha) \hspace{1mm} \hat{S}_2(\xi_E) \hspace{0.5 mm} \ket{0}\!\bra{0} \hspace{0.5mm} \hat{S}_2^\dag(\xi_E) \hspace{1mm} \hat{D}^\dag(\alpha),
\end{equation}

\noindent where $\hat{D}(\alpha) = \exp\left( \alpha \hat{a}^\dag - \alpha^\ast \hat{a} \right)$ is the displacement operator, $\hat{S}_2(\xi_E) = \exp\left( \frac{1}{2} \xi_E \hat{a}^\dag \hat{b}^\dag - \xi_E^\ast \hat{a}\hat{b} \right)$ is the two-mode squeeze operator (where $\hat{b}$ acts on Eve's idler mode) and $\ket{0}$ is the vacuum state. Without loss of generality we will let $\xi_E$ be real.

As is typical, we will model the loss due to the attenuator as a beam splitter. A fraction $\eta$ is allowed to pass through to reach Alice's apparatus, and $1-\eta$ is diverted into an auxiliary environment mode.

The final ingredient to be included is the thermal noise. This may be produced by heating up a portion of the optical fibre, so that Eve receives both her own photons that she sent in, as well as the thermal noise photons added by Alice. Here we need some careful thought as to how exactly we will \emph{mathematically} combine these two states. In other papers \cite{Lasota2017RobustnessNoise}, thermal noise has been added to a signal by passing both the signal and the noise through a beam splitter. However, this does not seem to us to be appropriate in this situation for the following reason. Suppose the combined state is produced by passing these two states through a beam splitter with transmissivity $\eta_T$, so that $\eta_T = 1$ means that the resulting state is entirely a thermal state, and $\eta_T = 0$ means it is all signal. However, this introduces a new variable into the situation, which implies some degree of coupling between the thermal source and Eve's returned state. We want Eve to be oblivious as to the actual source of the thermal noise, and simply consider it as a simultaneously arriving light source. In particular, if we let $\eta_T = 1$ and $\mu_T = 0$, we arrive at the rather paradoxical conclusion that the signal has been completely overwhelmed by a thermal state containing no photons. For a similar reason we cannot combine $\rho_\eta^\theta$ with a thermal density matrix $\rho_\textrm{Th}$ by way of a classical mixture such as $p \hspace{0.5mm} \rho_\eta^\theta + (1-p) \hspace{0.5mm} \rho_\textrm{Th}$. As such, we expect that the strength of the thermal noise should depend only on the single parameter $\mu_T$.

\begin{figure}[t]
   \centering
     \includegraphics[width=\columnwidth]{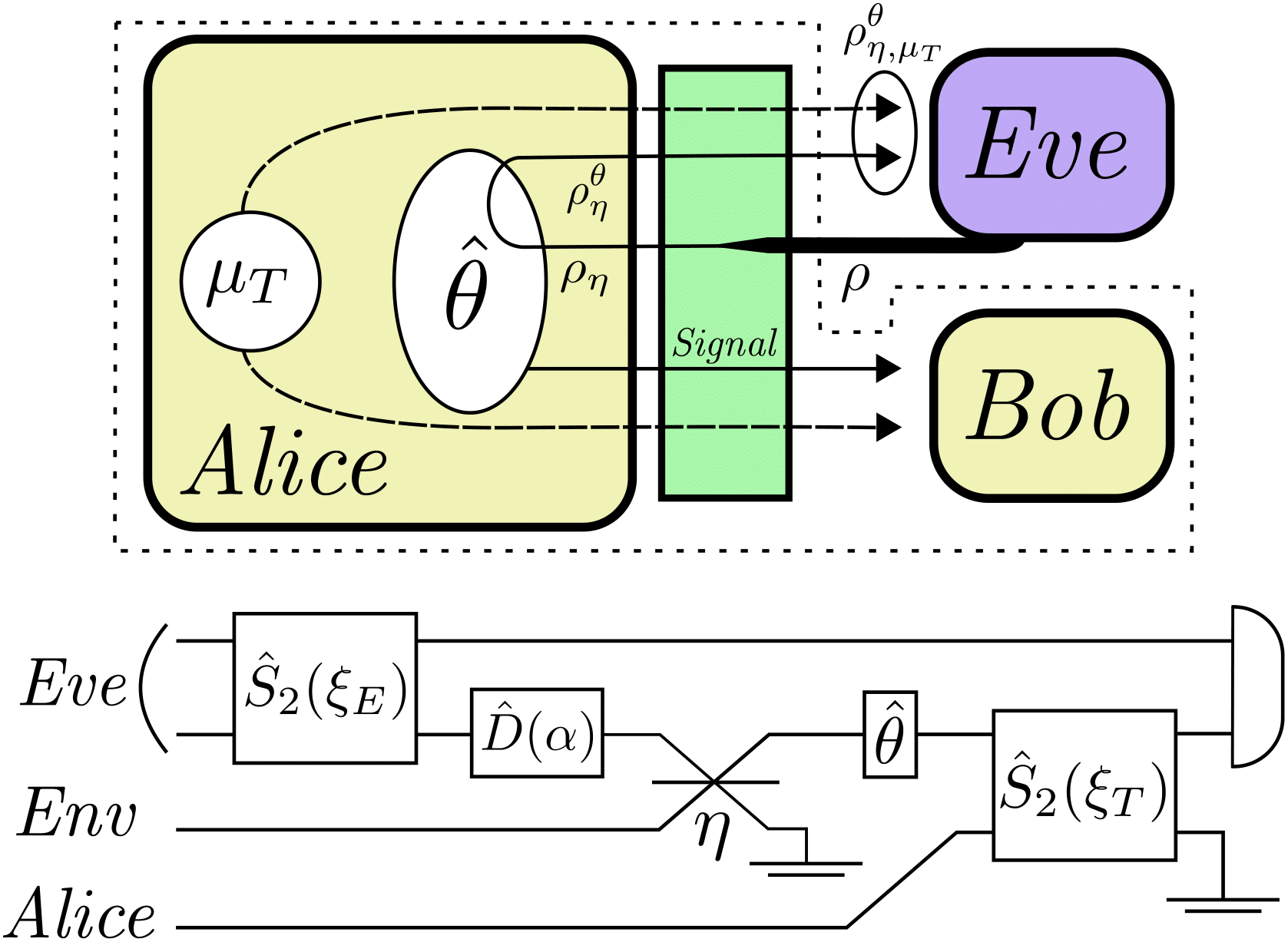}
     \captionsetup{justification=justified}
     \caption{Top: Schematic diagram illustrating the \emph{physical} mechanisms that produce Bob and Eve's states. The dotted line outlines the ``legitimate'' part of the protocol, comprising the signal and the thermal noise (dashed line). Bottom: Circuit diagram showing the \emph{mathematical} mechanisms that produce Eve's final state. Shown from left to right are the effects of Eve's squeezing, Eve's state displacement, Alice's attenuator, picking up the phase information, and adding the thermal noise. Double horizontal lines represent taking a partial trace over the relevant mode.}
     \label{fig:circuit}
\end{figure}

A method for the proper treatment of constructing a combined state from multiple simultaneously arriving photonic states was described by Glauber in his original treatment of the coherent states \cite{Glauber1963CoherentField}. However, that method involved expressing the states in a diagonal coherent basis (the so-called $P$-representation). Whilst this is a powerful method, it results in an expression for the state that is no longer easily analytically tractable (although it \emph{is} possible to use this to \emph{numerically} analyse the effects of adding non-thermal noise). Since we are dealing here with Gaussian states we shall take advantage of a nice property of thermal states: that they may be produced be taking the partial trace over one mode of a two-mode squeezed vacuum with squeezing parameter $\xi_T = \textrm{arcsinh}(\sqrt{\mu_T})$. Therefore, we shall model the addition of the thermal noise as the action of Alice passing Eve's returning signal through a two-mode squeezer with the vacuum, and discarding one of the resulting modes. Note that she does not \emph{physically} do this, it is only used to find the mathematical form of the state. Within this framework, Eve should choose $\alpha$ and $\xi_E$ to maximise her mutual information with the secret key. The full set-up for the construction of Eve's state is illustrated in Fig. \ref{fig:circuit}.

A great advantage of working with Gaussian states is that they may be completely characterised by their first and second moments. For an $n$-mode Gaussian state let $\hat{\underline{u}}$ be the vector of operators $\left[ \hat{x}_1 , \hat{p}_1 , \ldots , \hat{x}_n , \hat{p}_n \right]^T$. Then to each Gaussian state, $\rho$, we may uniquely assign a pair $\left(\underline{u} , V\right)$ which we call the \emph{mean vector} and \emph{covariance matrix} respectively, with elements defined by

\begin{equation} \label{eq:meanAnCovDef}
\begin{split}
u_i = \Tr\left[\rho\hat{u}_i\right]&, \\
V_{i,j} = \frac{\Tr\left[\rho\hat{u}_i \hat{u}_j\right]+\Tr\left[\rho\hat{u}_j \hat{u}_i\right]}{2}& - \Tr\left[\rho\hat{u}_i\right]\Tr\left[\rho\hat{u}_j\right].
\end{split}
\end{equation}

Let $\phi$ be the relative angle between the displacement and the squeezing parameters in the complex plane, $\mu_D = \eta\left|\alpha\right|^2$ be the average displacement after loss, and $\omega = \cosh(2\xi_E)$ be the normalised quadrature variance for a squeezed vacuum state. It then follows from Eq. \ref{eq:meanAnCovDef} and Eq. \ref{eq:defOfState} that the mean vector and covariance matrix for Eve's returned states corresponding to $\theta = 0$ and $\theta = \frac{\pi}{2}$ are as follows:

\begin{equation}\label{eq:covmats} 
\begin{split}
&u_{\theta=0} = 
\begin{bmatrix}
\left( \sin\phi+\cos\phi \right)\sqrt{2\mu_D} \\
\left( \sin\phi-\cos\phi \right)\sqrt{2\mu_D} \\
0 \\
0
\end{bmatrix},\\
&V_{\theta=0} = \frac{1}{2} 
\begin{bmatrix}
   \left[ \left(1+\mu_T\right)\omega \eta +\mu_T \right] \mathbb{1}_2 & A \hspace{1mm} \sigma_Z  \\[6pt]
  A \hspace{1mm}\sigma_Z & \omega \mathbb{1}
\end{bmatrix}, \\[6pt]
&u_{\theta=\frac{\pi}{2}} = 
\begin{bmatrix}
\left( \cos\phi-\sin\phi  \right)\sqrt{2\mu_D} \\
\left( \cos\phi+\sin\phi \right)\sqrt{2\mu_D} \\
0 \\
0
\end{bmatrix},\\
&V_{\theta=\frac{\pi}{2}} = \frac{1}{2} 
\begin{bmatrix}
   \left[ \left(1+\mu_T\right)\omega \eta +\mu_T \right] \mathbb{1}_2 & A\hspace{1mm} \sigma_X  \\[6pt]
  A\hspace{1mm} \sigma_X & \omega \mathbb{1}
\end{bmatrix},
\end{split}
\end{equation}

\noindent where $A = \sqrt{\left(1+\mu_T\right)\left(\omega^2-1\right)\eta}$ and $\sigma_Z$ and $\sigma_X$ are the Pauli $Z$ and $X$ matrices respectively. 

\subsection{Secret key rate} \label{subsec:secretkeyrate}

In all QKD systems the main quantity of merit is the secret key rate, $K$. This is the rate at which Alice and Bob can generate key bits with exponentially high security, which is in general lower than the rate at which Alice and Bob exchange raw key bits. This quantity is dependent on the specific choice of protocol that is being implemented. Here, we will analyse the performance of the BB84 protocol, which may be seen to be equivalent to entanglement-based protocols such as E91. This is because instead of deciding on a key bit $\ket{0_B}$ or $\ket{1_B}$ to send out for some basis $B$, Alice may instead prepare the entangled state $\left( \ket{0_B}\ket{\uparrow} + \ket{1_B}\ket{\downarrow} \right)/\sqrt{2}$, send the first mode to Bob and keep the second mode. She would then make a measurement of her retained mode to ``decide'' on the key bit. A similar process can be done to decide the basis. Since both frameworks are the same from the point of view of local density matrices as seen by Eve, the security of one reduces to the security of the other. 

In vanilla BB84 with no threat of THA, the secret key rate has been found to be \cite{gottesman_replaced}

\begin{equation} \label{eq:Kbasic}
K = R\left[ 1 - 2 H_2(\epsilon) \right],
\end{equation}

\noindent where $R$ is the raw key rate, $\epsilon$ is the bit-error rate and $H_2(\epsilon) = -\epsilon\log_2(\epsilon) - (1-\epsilon)\log_2(1-\epsilon)$ is the binary entropy function. We may say that one of these terms of $H_2(\epsilon)$ is due to Alice and Bob sacrificing key bits to perform error correction, and one factor is due to them applying classical privacy amplification algorithms.

Due to the nature of the THA being a SCA, Eve's attack will not affect the bit-error rate measured by Alice and Bob. However, it will still clearly compromise the security, so Eq. \ref{eq:Kbasic} cannot represent the achievable secret key rate. We expect in particular that the $H_2(\epsilon)$ term representing the error correction should remain unchanged --- since a properly implemented SCA will not induce additional errors. However, Alice and Bob \emph{will} have to do additional privacy-amplification, so this term will be modified.

The key rate for BB84 in the presence of an SCA was found by \cite{gottesman_replaced,Koashi2009SimpleComplementarity,Tamaki2003UnconditionallyStates}. They show that the effect of the SCA may be summarised by a quantity known as the \emph{distinguishability}, $\Delta$. This is used to modify the error rate, $\epsilon$, in the privacy-amplification term to become an \emph{effective error rate}, $\tilde{\epsilon}$ given by 

\begin{equation} \label{eq:tilde_epsilon}
\begin{split}
\tilde{\epsilon}(\epsilon,\Delta) =& \hspace{2mm} \epsilon + 4\Delta(1-\Delta)(1-2\epsilon) \\&+ 4(1-2\Delta)\sqrt{\Delta(1-\Delta)\epsilon(1-\epsilon)}.
\end{split}
\end{equation}

This means that we do not have to know exactly what Eve does with the states and the information available to her. For example, she may perform a THA to try to learn $\theta$ directly. Or, she might tailor her THA such that the measurement on the returned state only reveals information about the basis that Alice has chosen. After estimating this basis, she might then measure the \emph{flying} qubit in that basis to learn $\theta$ without disturbing the state. She might do some combination of these approaches, or something else entirely. As such, it is of foundational importance to our analysis that we have some way of quantifying the strength of a THA, that only makes reference to the state she sends \emph{out}, not to \emph{what she does} to the state she gets back, including any measurement or series of measurements on any combination of the returned state and flying qubits. The distinguishability varies from 0 when all choices of $\theta$ are indistinguishable from the point-of-view of Eve, and $\frac{1}{2}$ when she can distinguish all settings with certainty. In practice, a value of $\Delta$ much greater than 0 will result in a secret key rate of 0, since it would require Alice and Bob to be sacrificing raw key bits for error correction and privacy amplification at a rate faster than they are being generated. This formulation of the strength of a THA in terms of $\Delta$ puts a lower bound on the secret key rate that Alice and Bob can hope to achieve. The distinguishability is given by \footnote{This may be seen by considering Ref. \cite{gottesman_replaced}, section VIII. The purified states corresponding to each basis are as defined in Ref. \cite{Lucamarini2015PracticalDistribution}, Appendix B. From these it may be seen that $1-2\Delta$ is equal to the \emph{average}
 fidelity between a state being emitted in the $X$ basis and one in the $Y$ basis. By the symmetry and unitary invariance of the fidelity function, this reduces to finding the fidelity between only the states corresponding to $\theta = 0$ and $\theta = \pi/2$.}

\begin{equation} \label{eq:delta}
\Delta \leq \frac{1-F\left( \rho_{\eta,\mu_T}^0 , \rho_{\eta,\mu_T}^{\pi/2} \right)}{2},
\end{equation}

\noindent where $F\left( \rho_1 , \rho_2 \right) = \Tr\left[ \sqrt{\sqrt{\rho_1}\rho_2\sqrt{\rho_1}} \right]$ is the quantum fidelity function. Note that this is different from the form given in \citep{Lucamarini2015PracticalDistribution}. There, they reduce Eq. \ref{eq:delta} to a form involving the optimal purifications of the two output states. Since they are using pure coherent states, such optimal purifications are easily found. However, there exists no general prescriptive formula to find these for a pair of general mixed states, so we must use the fidelity form of the distinguishability. 

One may note that in this form, $\Delta$ has a nice physically intuitive interpretation. Suppose two states are prepared, one from the $X$ basis and one from the $Y$ basis, and one of them is given to Eve. She is aware of which two states are prepared but not which one she received. The quantity $\Delta$ then corresponds to the minimum probability that she makes an error in distinguishing them. If she succeeds in this task, she will know $\theta$, without needing to perform any additional operations on the flying qubits. This leads to the non-trivial conclusion that Eve's optimal THA may be performed by \emph{only} interacting with her own returned state, and she does not gain anything by interacting with Bob's qubits. 

The rest of this section is dedicated to calculating an exact expression for $\Delta$ for the set of thermalised Gaussian states described above, and section \ref{sec:separable} is focused on calculating a bound on $\Delta$ for the set of general separable states. It should be noted that, unlike $\epsilon$, $\tilde{\epsilon}$ (or equivalently $\Delta$) cannot be directly measured in the process of running the QKD protocol. Therefore Alice should be able to perform some local action to be able to determine $\Delta$ to some high precision, and then use this value to determine how much privacy amplification they should perform.

The problem of calculating the fidelity between two multimode Gaussian states was solved by \cite{Banchi2015QuantumStates}. There, they show that, for any Gaussian states $\rho_1, \rho_2$, we have:

\begin{equation}\label{eq:SPFidelity}
\begin{split}
F\left(\rho_1,\rho_2\right)&=\mathcal{F}\left(\tilde{V}_1,\tilde{V}_2\right) e^{-\frac{1}{4}\left(\underline{u}_1-\underline{u}_2\right)^T \left(\tilde{V}_1+\tilde{V}_2\right)^{-1}\left(\underline{u}_1-\underline{u}_2\right)}\\
\mathcal{F}\left(\tilde{V}_1,\tilde{V}_2\right) &= \frac{\prod_{k=1}^n\sqrt{w_k + \sqrt{w_k^2-1}}}{\sqrt[4]{\textrm{det}\left(\tilde{V}_1+\tilde{V}_2\right)}},
\end{split}
\end{equation}

\noindent where $\tilde{V}$ is equivalent to $V$, but expressed in the basis $\left[ \hat{x}_1 , \hat{x}_2 , \ldots , \hat{x}_n , \hat{p}_1 , \hat{p}_2 , \ldots , \hat{p}_n \right]^T$ and $w_k$ are the eigenvalues of the auxiliary matrix $W$, defined as

\begin{equation}\label{eq:auxilliary}
\begin{split}
W &= -2 i \Omega^T\left(\tilde{V}_1+\tilde{V}_2\right)^{-1}\left( \frac{\Omega}{4} +\tilde{V}_2\Omega \tilde{V}_1 \right)\Omega,\\
\Omega &= \begin{bmatrix}
   0 & 1  \\
  -1 & 0
\end{bmatrix} \otimes \mathbb{1}_n.
\end{split}
\end{equation}

When we combine the fidelity given in Eq. \ref{eq:SPFidelity} with the mean vectors and covariance matrices given in Eq. \ref{eq:covmats}, we find that the fidelity between two of Eve's returned states is given by 

\begin{equation}\label{eq:found_F}
\begin{split} %\tilde{F}(\mu_D,\omega,\eta,\mu_T) \equiv 
& F\left( \rho_{\eta,\mu_T}^0 , \rho_{\eta,\mu_T}^{\pi/2} \right) = \\ 
& \frac{1}{4B} e^{-2\mu_D\omega/B}\left(\sqrt{C} + \left| 4\mu_T\omega+4\eta(1+\mu_T)-1 \right|\right),
\end{split}
\end{equation}

\noindent where

\begin{equation}
\begin{split}
B = \hspace{1mm} &2\mu_T\omega + (1+\mu_T)(\omega^2+1)\eta, \\
C = \hspace{1mm} &16\eta^2(1+\mu_T)^2 + 8\eta(1+\mu_T)\omega(4\mu_T+\omega) \\&+ (1+4\mu_T\omega)^2.
\end{split}
\end{equation}

Eve wants to choose her parameters $\xi_E$ and $\mu_D$  in order to minimise the fidelity (and so maximise the distinguishability) between her returned states. Whilst increasing either of these parameters decreases $F$, she is not necessarily free to do both simultaneously. Both squeezing and displacement increase the average number of photons in each mode, and we have already established in subsection \ref{subsec:stateDescription} that this is limited by some number $N$.

Suppose, then, that Eve decides to use $pN$ of her available photons to contribute towards squeezing and $(1-p)N$ towards displacement. Since a squeezing parameter of $\xi_E$ gives an average photon number per mode of $\sinh^2\!\left(\xi_E\right)$, and a displacement parameter of $\alpha$ contributes $\left|\alpha\right|^2$ photons, we find that we can do no better that setting the parameters such that $\omega = \cosh\!\left[\textrm{arcsinh}\!\left(2\sqrt{pN}\right)\right], \hspace{1mm} \mu_D = (1-p)N\eta$ for some $p$.

When we insert these values into Eq. \ref{eq:found_F}, we can investigate the behaviour as a function of $p$ and $\eta$ for various values of $N$ and $\mu_T$. We find that $F$ is minimised when $p=0$. This means that Eve is best served by using \emph{all} of her photons to contribute to the displacement of her state. As such, we can now simplify the fidelity, which may be written as 

\begin{equation} \label{eq:simplifiedF}
F\left( \rho_{\eta,\mu_T}^0 , \rho_{\eta,\mu_T}^{\pi/2} \right) = \exp\!\left(-\frac{\mu_D}{1+2\mu_T}\right).
\end{equation}

This means that Eve's optimal Gaussian-state attack is one involving coherent states only. This provides a rigorous footing for earlier works which analyse the results of coherent-state attacks with an attenuating defense \cite{Lucamarini2015PracticalDistribution}.  

\subsection{Effect of thermal noise on Bob} \label{subsec:effectOnBob}

We can see from Eq. \ref{eq:simplifiedF} that Eve's knowledge of $\theta$ is minimised when $\mu_T$ becomes very large. However, when Alice sends a lot of thermal photons into the system, some of these photons are also picked up by Bob. If Bob measures these instead of the signal photons he is likely to pick up a bit error. Clearly when $\mu_T \rightarrow \infty$ Alice and Bob will not be able to securely communicate, so we need to find an optimal level of $\mu_T$ that clouds Eve's estimation of the state without affecting Bob too much. In this section we analyse and quantify this.

Consider the case where Alice and Bob are sending and receiving in the same basis. By implementing \emph{asymmetric BB84} \cite{Lo_replaced}, this can be the case for almost all qubits. We now say that at the end of the optical channel between Alice and Bob there is a polarising beam splitter which sends the incoming photons into one of two detectors which we label the \emph{correct} and \emph{wrong} detectors. These represent Bob measuring the bit that Alice did and did not send respectively. 

Alice and Bob will try to distill a secret key from the key bits where Bob believes he detected only a single photon (which he must assume to be the signal photon). The probability for this to occur for a given signal qubit we will call $p_\textrm{succ}$. Here we will assume that Bob uses bucket detectors. That is to say they have two measurement outcomes: either no photons were detected, or one or more photons were detected. We show in Appendix \ref{app:numberResolving} that, perhaps counter-intuitively, this actually gives a \emph{better} secret key rate than using number-resolving detectors, in agreement with previous results \cite{Lasota2017RobustnessNoise}.

Let $p_\checkmark, p_\times$ be the probabilities that the signal photon was detected in the correct and wrong detector respectively, and $p_\bullet$ be the probability that the signal is not detected at all. Let $q(c,w)$ be the probability that $c$ noise photons are detected in the correct detector and $w$ in the wrong detector. Bob will register a ``valid'' qubit if exactly one of the detectors clicks. We can then say that 

\begin{equation} \label{eq:p_succ}
\begin{split}
p_\textrm{succ} =& \hspace{1mm} p_\checkmark \sum_{c=0}^\infty q(c,0) + p_\times \sum_{w=0}^\infty q(0,w)\\
				+& \hspace{1mm} p_\bullet \sum_{c=1}^\infty q(c,0)  + p_\bullet \sum_{w=1}^\infty q(0,w). 
\end{split}
\end{equation}

We can identify the bit error rate, $\epsilon$, as the probability that a photon gives a click in the wrong detector, and is therefore equal to 

\begin{equation}
\epsilon = \hspace{1mm} \frac{p_\times \sum_{w=0}^\infty q(0,w)
				+ p_\bullet \sum_{w=1}^\infty q(0,w)}{p_\textrm{succ}}.
\end{equation}

To calculate these quantities, we identify the following:

\begin{equation}
\begin{split}
p_\checkmark &= T(1-Q), \\
p_\times &= TQ \\
p_\bullet &= (1-T) \\
q(i,j) &= \frac{\tilde{\mu_T}^i}{(\tilde{\mu_T}+1)^{i+1}} \frac{\tilde{\mu_T}^j}{(\tilde{\mu_T}+1)^{j+1}} \\
\sum_{c=0}^\infty q(c,0) &= \sum_{w=0}^\infty q(0,w) = \frac{1}{\tilde{\mu_T}+1} \\
\sum_{c=1}^\infty q(c,0) &= \sum_{w=1}^\infty q(0,w) = \frac{\mu_T}{(\tilde{\mu_T}+1)^2}.
\end{split}
\end{equation}

Here, $Q$ is the probability for Bob to register a bit-flip error in the absence of thermal noise, $T$ is the transmissivity of the channel including the efficiency of Bob's detectors, and $\tilde{\mu_T} = T\mu_T/2$ is the average number of thermal photons arriving at each detector. In practice, $\tilde{\mu_T}$ may be lower than this, since Eve will have inadvertently intercepted some of them. However, we give here the worst-case scenario. We can then say that 

\begin{equation}
\begin{split}
p_\textrm{succ} &= \frac{2T(2+2\mu_T-T\mu_T)}{(2+T\mu_T)^2},\\[6pt]
\epsilon &= \frac{2Q+\mu_T(1-T+QT)}{2+\mu_T(2-T)}.
\end{split}
\end{equation}

In order to find an expression for $T$ we assume that the photons face an exponential drop-off with distance, and set $T=e^{-L/L_0}$ where $L_0$ is the attenuation length. To model $Q$, we may assume that, as is usual for QKD protocols, Alice and Bob are equipped with quantum memories, and the flying qubits are used as a process by which they establish entanglement between these memories \cite{Simon2007QuantumMemories,nemoto_replaced,Zwerger2012Measurement-basedRepeaters,Vinay2017PracticalCommunication}. This is necessary for all but the most primitive protocols, since some storing of entangled qubits is required in order to perform entanglement distillation and privacy amplification algorithms such as DJEMPS \cite{Deutsch1996QuantumChannels}, which are needed in order to prove that Eve has not entangled Bob's state with an ancilla. As such, we set $Q=\frac{1}{2}\left[1-e^{Lc/\tau}\right]$, where $\tau$ is the lifetime of the memory (typically on the order of microseconds). Finally, when $p_\textrm{succ}<1$, we must replace $\Delta$ with $\Delta/p_\textrm{succ}$, since the lost signals may have been selectively eliminated by Eve to improve her mutual information with the key (Ref. \cite{gottesman_replaced}, Eq. 32). 

\subsection{Results}

By combining these elements, which measure the effects of the thermal noise on Eve and on Bob, as well as the result that coherent states are optimal, we find that the final secret key rate for a general multi-mode Gaussian state attack in the presence of an attenuating filter and a thermal noise defense may given by 

\begin{equation} \label{eq:final_thermal_keyrate}
K=p_\textrm{succ}\left[1-H_2\left(\epsilon\right)-H_2\left(\tilde{\epsilon}\left(\epsilon,\Delta'\right)\right)\right],
\end{equation}
\noindent where $\Delta'=\left[1-\exp(-\frac{\mu_D}{1+2\mu_T})\right]/\left(2 \hspace{0.5mm}p_\textrm{succ}\right)$.

\begin{figure}[t]
   \centering
     \includegraphics[width=\columnwidth]{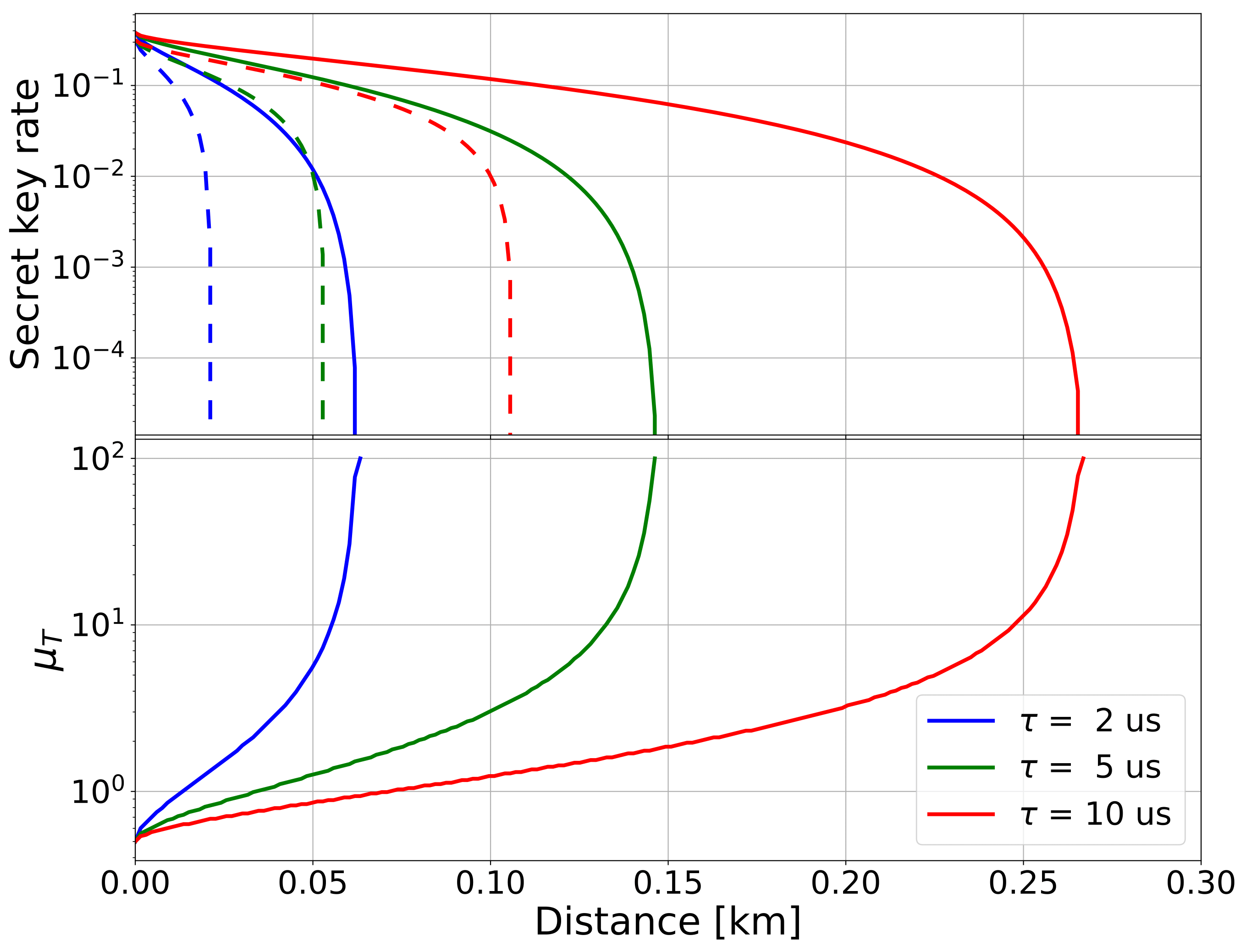}
     \captionsetup{justification=justified}
     \caption{Minimum achievable secret key rate under the presence of a coherent state attack (the most powerful Gaussian-state attack), with a thermal noise defense involving $\mu_T$ photons being added by Alice to cloud Eve's attack signal. Top graph shows the key rate for different values of quantum memory time, $\tau$. Dashed lines show the key rate with no thermal noise added, and solid lines are key rates maximised over  $\mu_T$. From left to right for both dashed and solid lines, we have $\tau = 2, 5, 10 \hspace{0.5mm} \mu s$ respectively. Lower graph shows the optimal values of $\mu_T$ at each distance. In all plots $\mu_D = 0.1$ and $L_0 = 25 \hspace{0.5mm} \textrm{km}$.}
     \label{fig:SKR_main}
\end{figure}

Eq. \ref{eq:final_thermal_keyrate} is highly dependent on $\mu_T$. We optimise $K$ over $\mu_T$ to find a true measure of the utility of the thermal noise defense. We consider the case where $\mu_D=0.1$ and $L_0=25 km$. The ability of the thermal noise defense to protect the secret key rate is shown in Fig. \ref{fig:SKR_main}.

We can see from this that employing a thermal noise defense has the capacity to more than \emph{double} the effective range of a QKD system. However, when the key rates get very small, very high temperatures are required in order to retain any security. This is because the mechanism by which increasing $\mu_T$ decreases the key rate is by decreasing the pre-factor $p_\textrm{succ}$, which may go arbitrarily low, but never reach 0. On the other hand, there is a small \emph{critical} effective error-rate not equal to 1 that Eve can induce that will result in the key rate dropping to zero. Therefore, when Alice calculates that the effective error that could result from Eve launching a THA is above the critical value, Alice is forced to increase the thermal noise to very high values to bring the effective error down again. This mechanism of sacrificing raw key rate is common amongst QKD protocols, and is the general basis for privacy amplification \cite{Deutsch1996QuantumChannels,renner2005universally}. 

One may notice that the ranges shown in Fig. \ref{fig:SKR_main} are well below those shown in many other QKD proposals. One reason for this is that other proposals in general do not consider the effects of THAs, or SCAs in general, and the key-rate boon provided by the thermal noise defense cannot exceed the calculated key rate of a system that does not recognise SCAs in the first place. Secondarily, the key-rate plots shown here are intended to show the \emph{relative advantage} of adding thermal noise to the system. They are not intended to show the state-of-the-art ranges that might be achieved by more advanced QKD protocols, which may take advantage of technologies such as quantum repeaters \cite{Vinay2017PracticalCommunication,Duan2001Long-distanceOptics.,Piparo2015Long-DistanceDistribution,epping_replaced}, long-lived quantum memories \cite{piparo_replaced,nemoto_replaced} or post-selective entanglement generation \cite{Kok2007LinearQubits,Vinay2017PracticalCommunication}. 

\section{General Separable Attacks} \label{sec:separable}

We have shown that, amongst multi-mode Gaussian attack states that Eve might use, the separable coherent state is optimal. Whilst it may seem initially surprising that entanglement does not assist her, note that entanglement between any two modes will drop off as more of the signal is attenuated. We are left with a distinguishability that depends only on the average output photon number, $\mu_D$, so $\Delta$ does not \emph{explicitly} depend on the transmissivity $\eta$.

It may be argued that coherent states are likely to be optimal amongst the separable states, since under loss, photon-counting statistics will tend to be Poissonian \cite{Hu2007OnLasers}. Therefore the best one can hope to do is with a Poissonian state that retains coherence, i.e. a coherent state. However, a state that is initially highly non-Poissonian in its statistics may require a very high attenuation before it approximates a Poissonian distribution, and there is no guarantee that the expression for $\Delta$ derived from coherent states will still hold.

In this section, we consider the set-up where Alice defends against a THA by use of an attenuator but uses no thermal defense, and that Eve attacks the system using \emph{any} separable state, but gets back a state with only a few photons. Whilst this seems to be a special case for Eve, note that it is more general that the situation considered in section \ref{sec:thermalnoise} since this approach considers a set of states which includes, yet is larger than, the set of coherent states that are optimal within the Gaussian states.

Here, we will consider Eve's input state in its density matrix form instead of its covariance matrix form. We will consider the effect of the attenuator on $\rho$ as a quantum channel, which we express in Kraus operator form:

\begin{equation}
\begin{split}
&\mathcal{E}\left(\rho\right) = \sum_{k=0}^\infty \mathcal{E}_k \left(\rho\right) \equiv \sum_{k=0}^\infty\hat{A}_k\rho\hat{A}_k^\dag \\
\hat{A}_k& = \sum_{j=k}^\infty \sqrt{j\choose k} \sqrt{\eta}^{\hspace{0.5mm}j-k}\sqrt{1-\eta}^{\hspace{0.5mm}k} \ket{j-k}\!\bra{j}.
\end{split}
\end{equation}

Each term $\hat{A}_k\rho\hat{A}_k^\dag$ represents $k$ photons being lost from the state $\rho$, each with independent probability $\eta$.

We express each term in the map as follows

\begin{equation}
\begin{split}
\mathcal{E}_k\left(\rho\right) &= \sum_{i,j=k}^\infty B_{i,j,k}(\eta) \bra{i}\rho\ket{j} \cdot \ket{i-k}\!\bra{j-k} \\
B_{i,j,k}(\eta) &= \sqrt{{i\choose k}{j\choose k}} \eta^{\frac{i+j}{2}-k} (1-\eta)^k.
\end{split}
\end{equation}

Since we require a very high level of attenuation to achieve any kind of useful secrecy, we may assume that $\eta$ is very close to 0. Therefore we may expand the factor $B_{i,j,k}(\eta)$ as 

\begin{equation}
B_{i,j,k}(\eta) \approx B_{i,j,k}(0) + \frac{\textrm{d}B}{\textrm{d}\eta}\Bigr|_0 \eta + \frac{1}{2} \frac{\textrm{d}^2B}{\textrm{d}\eta^2}\Bigr|_0 \eta^2,
\end{equation}

\noindent leading to an expansion of each term in the map as 

\begin{equation}
\mathcal{E}_k \approx \mathcal{E}_k^{(0)} + \mathcal{E}_k^{(1)} + \mathcal{E}_k^{(2)}.
\end{equation}

Using the fact that $\lim_{x\rightarrow 0}x^p = \delta_{p,0}$ for $p\geq 0$, we can see that $B_{i,j,k}(0) = \sqrt{{i\choose k}{j\choose k}} \delta_{\frac{i+j}{2},k}$. Performing the sums over $i,j,k$ we get $\mathcal{E}^{(0)} = \ket{0}\!\bra{0}\sum_{k=0}^\infty\bra{k}\rho\ket{k} = \ket{0}\!\bra{0}$ (where we have used the fact that $\Tr\left[\rho\right]=1$)

In the same way, we find that

\begin{equation}
\begin{split}
&\mathcal{E}_k^{(1)}=\sum_{i,j=k}^\infty \sqrt{{i\choose k}{j\choose k}} \Bigg[ \left(\frac{i+j}{2}-k\right)\underbracket[0.5pt][7pt]{\delta_{\frac{i+j}{2},k+1} }_{\mathclap{\delta_{i,k}\delta_{j,k+2}+\delta_{i,k+2}\delta_{j,k}+\delta_{i,k+1}\delta_{j,k+1}}}- k\delta_{\frac{i+j}{2},k}\Bigg]\eta\\
\vspace{5mm}&\therefore\hspace{2mm}\mathcal{E}^{(1)}=-\mu\ket{0}\!\bra{0} + \\
&\hspace{22mm}\mu \ket{1}\!\bra{1} + \\
&\hspace{20mm}\sum_{k=0}^\infty \sqrt{{(k+2)}\choose k}\ket{0}\!\bra{2}\bra{k}\rho\ket{k+2} + \\
&\hspace{20mm}\sum_{k=0}^\infty \sqrt{{(k+2)}\choose k}\ket{2}\!\bra{0}\bra{k+2}\rho\ket{k},
\end{split}
\end{equation}

\noindent where $\mu = \eta\sum_{k=0}^\infty k \bra{k}\rho\ket{k}$ is the average number of photons that Eve receives back after attenuation.

Similarly,

\begin{equation}
\begin{split}
\mathcal{E}^{(2)}=&\frac{\eta^2}{2}\left(v+\left\langle\hat{n}\right\rangle_{\rho}^2+\left\langle\hat{n}\right\rangle_{\rho}\right)\ket{0}\!\bra{0} - \\
&\eta^2 \left(v+\left\langle\hat{n}\right\rangle_{\rho}^2+\left\langle\hat{n}\right\rangle_{\rho}\right) \ket{1}\!\bra{1} + \\
&\frac{\eta^2}{2}\left(v+\left\langle\hat{n}\right\rangle_{\rho}^2+\left\langle\hat{n}\right\rangle_{\rho}\right)\ket{2}\!\bra{2}+\\
\vspace{2mm}&\textrm{off-diagonals on }\hspace{1mm}\ket{0}\!\bra{2}\hspace{1mm}\textrm{and}\textbf{}\ket{2}\!\bra{0} + \\
\vspace{2mm}&\textrm{terms on} \hspace{1mm} \ket{i}\!\bra{j} \hspace{1mm} \textrm{where} \hspace{1mm} i \hspace{1mm} \textrm{or} \hspace{1mm} j \geq 3.
\end{split}
\end{equation}

\noindent where $v=\left\langle\smash{\hat{n}^2}\right\rangle_{\rho}-\left\langle\smash{\hat{n}}\right\rangle_{\rho}$ is the variance in the initial state.
Whilst the diagonal terms can be expressed in terms of the macroscopic observables of $\rho$, the off-diagonal terms have no such simple expression. 

Firstly we should bound the effects of higher-order terms. We do this by supposing that Eve performs a measurement on her returned state to determine whether or not the state contains 2 or fewer photons. That is to say, her measurement of $\rho$ has 2 outcomes corresponding to operators $\hat{E}_\checkmark = \ket{0}\!\bra{0}+\ket{1}\!\bra{1}+\ket{2}\!\bra{2}$ and $\hat{E}_\times = \sum_{k=3}^\infty\ket{k}\!\bra{k}$.

In order to ensure that this measurement does not reduce the information that Eve learns about the state, we say that if she gets the result corresponding to $\hat{E}_\times$ then we assume that she learns the key bit $\theta$ perfectly. That is to say, instead of receiving $\mathcal{E}\left(\rho\right)$ she can be said to receive some state $\ket{\theta}\!\bra{\theta}$, where $\bra{\theta_1}\ket{\theta_2} = \delta_{\theta_1,\theta_2}$.

If the measurement is successful, then Eve's state is projected onto the two-or-fewer-photon subspace, giving $\rho_\textrm{sub} = \hat{E}_\checkmark\mathcal{E}(\rho)\hat{E}_\checkmark/\Tr\left[\hat{E}_\checkmark\mathcal{E}(\rho)\right]$. In the case where $\theta=0$, this may be expressed in the basis of $\left\{ \ket{0}, \ket{1}, \ket{2} \right\}$ by 

\begin{equation} \label{eq:rho_subspace}
\rho_\textrm{sub}^{\theta=0} = 
\begin{bmatrix}
   1-\mu+\frac{\eta^2}{2}\alpha & 0 & \beta  \\[6pt]
   0 & \mu-\eta^2\alpha & 0 \\[6pt]
   \beta & 0 & \frac{\eta^2}{2}\alpha
\end{bmatrix},
\end{equation}

\noindent where $\alpha = v+\left\langle\hat{n}\right\rangle_{\rho}^2+\left\langle\hat{n}\right\rangle_{\rho}$ and $\beta$ is some coefficient that cannot be easily expressed in terms of macroscopic properties of the state. In the case where $\theta = \pi/2$ we simply pick up a factor of $-1$ on the coefficient $\beta$. The overall state Eve gets back is then

\begin{equation} \label{eq:rho_returned_separable}
\rho_\textrm{returned}^\theta = \Tr\left[\hat{E}_\checkmark\mathcal{E}(\rho)\right]\rho_\textrm{sub}^{\theta} + \Tr\left[\hat{E}_\times\mathcal{E}(\rho)\right]\ket{\theta}\!\bra{\theta}.
\end{equation}

In order to bound the contribution of the second term, we show in Appendix \ref{app:proofOfEq} that

\begin{equation} \label{eq:EtickTrace}
\Tr\left[\hat{E}_\checkmark\mathcal{E}(\rho)\right] \geq e^{-\mu},
\end{equation}

\noindent and so $\Tr\left[\hat{E}_\times\mathcal{E}(\rho)\right] \leq 1-e^{-\mu}$.

Since we want to find an upper limit to the information that Eve can learn, we say that she can receive any state that is consistent with both Eq. \ref{eq:rho_subspace} and the laws of physics. We find that the fidelity between two such density matrices is minimised when the variance $v$ is chosen such that the $\ket{1}\!\bra{1}$ component is 0 and the off-diagonal terms are maximised. Because of this, it turns out that we do not need to be able to express $\beta$ in a way relating to macroscopic properties such as average photon number and variance. We simply choose $\beta$ to be the largest value such that $\rho_\textrm{sub}^{\theta}$ remains positive semi-definite, which is $\beta = \frac{1}{2}\sqrt{\mu(2-\mu)}$.

This, along with Eq. \ref{eq:rho_returned_separable} and Eq. \ref{eq:delta} gives an ultimate distinguishability bound for the separable attack-state case of 

\begin{equation} \label{eq:separable_distinguishability_bound}
\Delta \leq \frac{1-e^{-\mu}\sqrt{1-3\mu(2-\mu)/4}}{2}.
\end{equation}

\begin{figure}[t]
   \centering
     \includegraphics[width=\columnwidth]{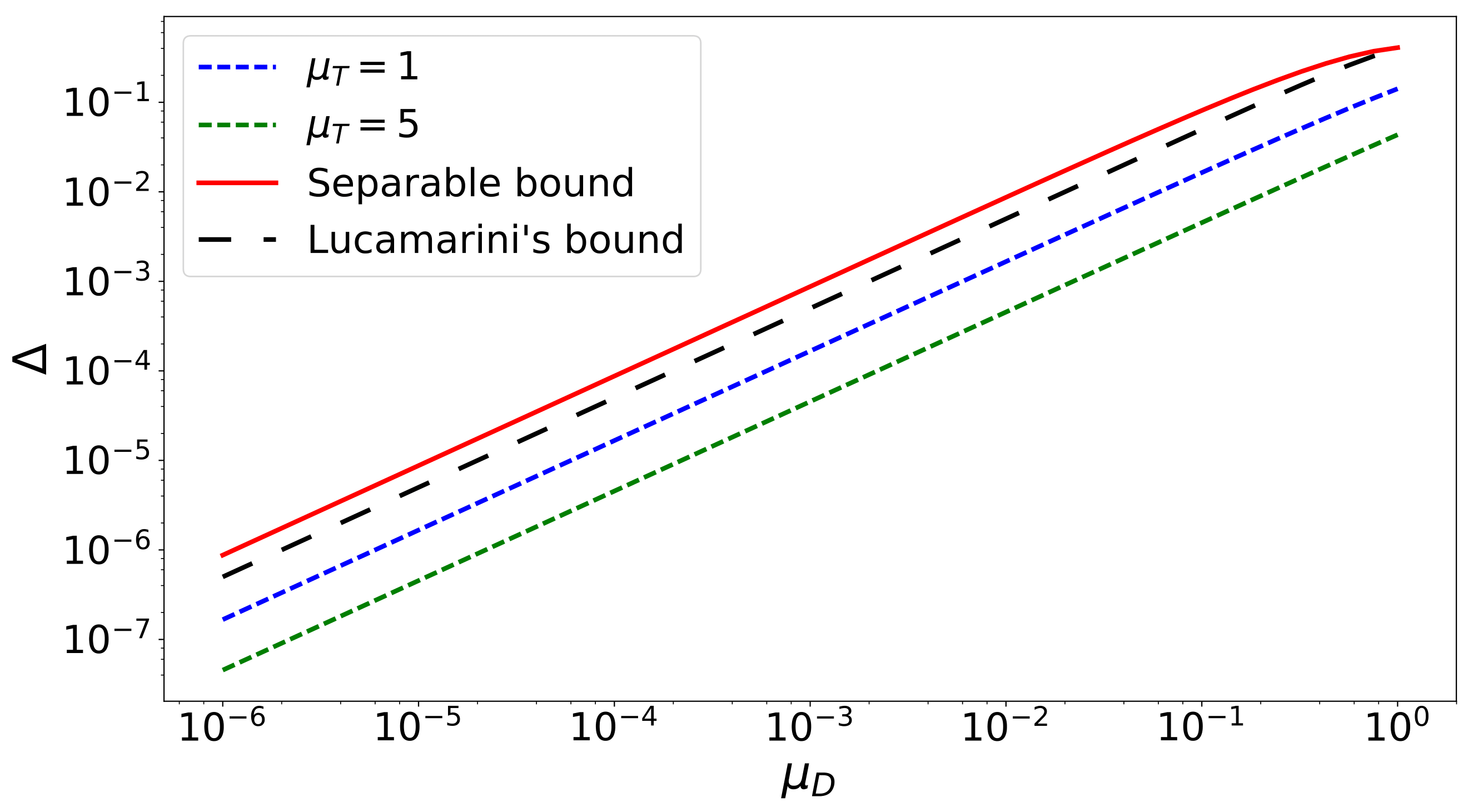}
     \captionsetup{justification=justified}
     \caption{Upper bounds on the distinguishability, with average returned photon number $\mu$. Dotted lines show the distinguishability for a coherent-state attack with a thermal noise defense, of thermal photon number $\mu_T$. Upper dotted line (blue) shows $\mu_T = 1$, lower dotted line (green) shows $\mu_T = 5$. The black dashed line is the distinguishability for coherent-state attacks found by Lucamarini et. al. Solid line is the bound for separable states, which as expected is always greater that the other bounds.}
     \label{fig:deltaCompare}
\end{figure}

Importantly, this is a function of a single variable, that is \emph{measurable by Alice}; the average output photon number. By bounding quantities of Eve's state that cannot be measured, we have ensured that Alice can make an accurate assessment of how secure her QKD system is, whilst not knowing anything about the microscopic details of Eve's state. 

Fig. \ref{fig:deltaCompare} shows that the value of $\Delta$ is higher for our separable bound than for the case of a coherent state, whether one diluted by thermal noise or not. We also show it to be higher than the bound on $\Delta$ for a noiseless coherent-state attack found by Lucamarini et al. of $\Delta = [1-e^{\mu}\cos(\mu)]/2$. Whilst our bound on $\Delta$ is not absolutely tight (since the 3-photon contributions surely will not convey perfect information of $\theta$), we can see that is not too generous, since it tracks the known achievable bounds quite closely.

\section{Shutter Defense} \label{sec:shutter}

In both sections \ref{sec:thermalnoise} and \ref{sec:separable}, it was necessary that Alice use a strong one-way attenuator. This must be able to let almost all of the light through in a forward direction, whilst blocking all but perhaps a single photon going in the reverse. Whilst such devices certainly exist \cite{Jalas2013WhatIsolator,Fujita2000WaveguideInterferometer}, achieving such a high level of attenuation can be a great technical challenge. We should ask if there is any other device we can insert into the optical channel that will result in a large \emph{effective} attenuation factor.

We show here that this is possible by considering the effect of replacing the attenuator with a \emph{shutter}, such as an electro-optic modulator or a chopper. This runs contrary to the claim of \cite{Gisin2002QuantumCryptography} Sec. VI K, where it is claimed that a shutter cannot defend against a THA (although they consider only a shutter directly adjacent to the apparatus, thus setting to travel time for the light to zero). This lets light through for a short duration of $t_S$, at a period of $t_P$. Once Eve's pulse passes the shutter, it must travel the remaining distance to the apparatus that encodes $\theta$, reflect off it, and return to the shutter. If the shutter is closed when the pulse arrives, it will reflect off the rear-side of the shutter with some co-efficient of reflectivity $\eta_R$, return to the apparatus, and then reflect again. This will continue until the pulse of light arrives back at the shutter whilst it is open, and it will then pass through and be detected by Eve. Note that whilst the pulse will pick up an additional phase factor of $\theta$ on each reflection, we will hold by the principle of assuming that Eve's computational and measurement power is the maximal allowed by the laws of physics. As such, it is plausible that she will be able to know by the time taken for the pulse to return exactly how many times the light reflected off the apparatus, and so calculate an estimate for the actual value of $\theta$ from her measured value.
 
Let $t_L$ be the time period that a pulse of light takes to make the return journey from the shutter to the encoding apparatus and back. If the light makes $R$ return journeys, then it will have reflected off the rear-side of the shutter $R-1$ times. One may see that $R$ can be found to be the smallest integer such that 

\begin{equation}
0 \leq \left( R \times t_L \hspace{1mm} \textrm{mod} \hspace{1mm} t_P \right) \leq t_S.
\end{equation}

Note that if the light travel time is known to arbitrary precision and $t_S$ can be made arbitrarily short, then $R$ can be made arbitrarily high, by letting $t_L = t_P - \delta$ for some arbitrarily small $\delta$. This, however, is physically unrealistic. We should instead model the shutter as being open for some finite fraction of the light travel time. Suppose initially that the shutter is open for a tenth of the period, i.e. $t_S = t_P/10$. We may say that the light travel time may be varied by implementing various lengths of coiled optical fibre between the shutter and the apparatus.

\begin{figure}[t]
   \centering
     \includegraphics[width=\columnwidth]{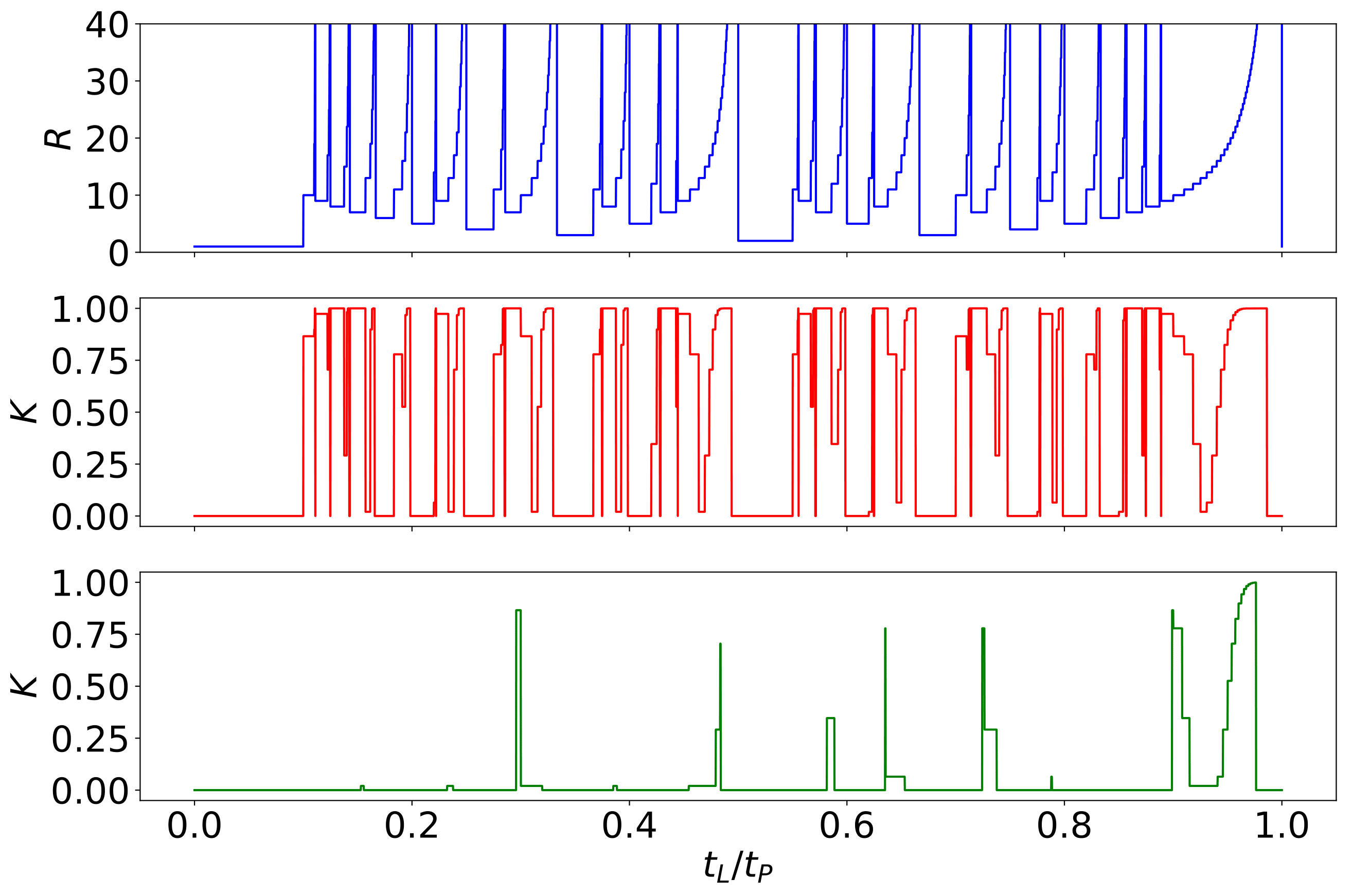}
     \captionsetup{justification=justified}
     \caption{Top: The number of reflections a pulse of light will make, as a function of the light travel time between the shutter and the apparatus. Middle: The resulting secret key rate. Here, we have used $\eta_R = 0.5$, meaning half of the light is lost upon each reflection. Bottom: The minimum secret key rate that can be guaranteed if there is a 1\% margin of error in the knowledge of the light travel time, found using a minimising convolution functional.}
     \label{fig:shutterDef3}
\end{figure}

The upper part of Fig. \ref{fig:shutterDef3} shows the values of $R$ that result from varying light travel times, where the light travel time is measured in units of $t_P$. In the middle part of Fig. \ref{fig:shutterDef3} we convert this value of $R$ into a secret key rate. We do this by using Eq. \ref{eq:final_thermal_keyrate} to find $K$ (with $p_{succ} = 1$) and use the separable bound Eq. \ref{eq:separable_distinguishability_bound} for $\Delta$. We may say that the average output photon number is given by $\mu = N\hspace{.5mm}\eta_R^{R-1}$, where $N$ is the input photon number.

The lower part of Fig. \ref{fig:shutterDef3} is the achievable key rate after the application of what we call a \emph{minimising convolution}. This is a functional which takes a function $f(\cdot)$, and maps each point $x$ to the minimum of $\{f(x+x') \hspace{1mm} | \hspace{1mm} x'\! \in\! [-\delta,\delta]\}$ for some convolution width $\delta$. This is necessary because there may be some experimental uncertainty in the light travel time. So, for example, whilst a value of $t_L$ infinitesimally close to, but less than $t_P$ may seem to give the highest value of $R$, and so the highest key rate, if $t_L$ was even slightly underestimated, this would result in a value of $R=1$ and so a far lower security would be achieved.
For Fig. \ref{fig:shutterDef3} we have used $\delta = 0.01$. This means that if we have a 95\% confidence interval of knowing the light travel time to within 1\%, then we can have a confidence of 95\% of being able to achieve the key rate shown in the lower graph.

Note that if we fix the light travel time appropriately (to approximately about $0.9 \times t_P$) then we can achieve a secret key rate of almost unity from a co-efficient of reflectivity of $\eta_R = 0.5$. One thing that one should bear in mind is that this is strongly dependent on the width of the confidence interval on $t_L$. Since we take the minimum of the interval, it is clear that if the interval is too wide, then no secret key rate will be able to be guaranteed. 

If one wishes to halve the width of the relative confidence interval, this can be done by doubling both the light travel time and the shutter period. However, this will result in a longer time between raw key bit attempts, which will have a lowering effect on the key rate. Similarly, if one tries to increase the key rate whilst using a shutter defense, one should be mindful of the effect it has on the minimum achievable key rate. For example, if the raw key rate was doubled, the uncertainty in $t_L$ could increase from 1\% to 2\%. We find that this results in a maximum key rate after the application of a minimising convolution of around 0.75, and so increasing the overall secret key rate to $2\times0.75=1.5$ of the original value. Similar to the case with adding thermal noise, in any experimental realisation one will have to adjust the light travel time to find the right balance between increasing raw key rate and increasing secrecy. In any case, this brief analysis shows that the use of a shutter as a defense against the THA is one worthy of consideration, and may provide effective attenuations comparable to those of directional attenuators. The choice as to which to use in any given implementation will depend on the details of the experimental set-up.

\section{Conclusion}

The discovery and implementation of the Trojan Horse Attack once threatened to eliminate the security so famously promised by quantum key distribution. Early seminal works have shown that the situation is not hopeless, and have indicated ways to quantify and abate this threat. 

In this work we have fully characterised and quantified the effect of the THA on the key rate under two general attack vectors. We have shown that if Eve uses a multimode Gaussian attack state, her best bet is to use a coherent state. We have also quantified the maximum damage on the secrecy that could be caused by Eve using any separable state. We hope that this may be extended to the general entangled case in the future, but we have provided heuristic arguments for why we do not expect much of an improvement for Eve by doing this. 

We have described two novel ways of counteracting the THA; a passive defense, enabled by adding thermal noise into the system, and an active defense with an optical modulator. These complement the attenuation-based defense discussed in earlier work.

This all shows that side-channel attacks cannot be considered only as an afterthought in QKD systems. Even a relatively rudimentary SCA can, if not protected against, hugely reduce the security of a protocol. If we try to improve the security by privacy amplification alone we find that the secret key rate soon drops to zero. This highlights the importance of proper and specific defenses against SCAs that are easily quantified in terms of experimentally accessible quantities.

\section*{Acknowledgments}
\noindent We wish to acknowledge the contributions of valuable discussions with \mbox{Mark} \mbox{Pearce} and \mbox{Stefano} \mbox{Pirandola}.

\noindent\rule{\columnwidth}{0.4pt}

\appendix

\section{Use of number-resolving detectors} \label{app:numberResolving}

Our calculation of the secret-key rate is based on the fact that Bob uses bucket detectors. One might naturally ask whether using more state-of-the-art technology such as photon-number-resolving detectors (PNRDs) would improve the situation for Bob. 

When this is the case, Eq. \ref{eq:p_succ} becomes

\begin{equation} \label{p_succ}
\begin{split}
p_\textrm{succ} =& \hspace{1mm} p_\checkmark\hspace{1mm} q(0,0) + p_\times\hspace{1mm} q(0,0)\\
				+& \hspace{1mm} p_\bullet\hspace{1mm} q(1,0)  + p_\bullet\hspace{1mm} q(0,1), \\
                \vspace{6mm}
                =& \hspace{1mm}T\hspace{1mm}\frac{1}{(\tilde{\mu_T}+1)^2} + 2\hspace{1mm}(1-T)\hspace{1mm}\frac{1}{\tilde{\mu_T}+1}\frac{\tilde{\mu_T}}{(\tilde{\mu_T}+1)^2},
\end{split}
\end{equation}

\noindent since a click is only registered when \emph{exactly} one photon enters a detector.

When we calculate $\epsilon$ by taking the sum of the contributions to $p_\textrm{succ}$ that cause the \emph{wrong} detector to click conditioned on the probability of getting a click in the first place, we find that $\epsilon$ is actually the same whether we use PNRDs or bucket detectors! This interesting congruence is a result of the fact that the noise obeys thermal statistics, and will not generally be true for other noise distributions. However, whilst the relative error is not affected, $p_\textrm{succ}$ is actually \emph{lower} in the case of PNRDs! This means that the secret key rate can not be improved by using PNRDs, and is in fact worsened in almost all cases. This is because the restriction to PNRDs means that Bob is more likely to reject legitimate signals than noise photons.

\section{Proof of Eq. \ref{eq:EtickTrace}} \label{app:proofOfEq}

We want to show that $\Tr\left[\hat{E}_\checkmark\mathcal{E}(\rho)\right] \geq e^{-\mu}$.

First note that since $\hat{E}_\checkmark = \ket{0}\!\bra{0}+\ket{1}\!\bra{1}+\ket{2}\!\bra{2}$, we can assert that 

\begin{equation} \label{eq:app:firsteq}
\Tr\left[\hat{E}_\checkmark\mathcal{E}(\rho)\right] \geq \Tr\left[\ket{0}\!\bra{0}\mathcal{E}(\rho)\right].
\end{equation}

Since $\mathcal{E}$ does not map off-diagonal elements to diagonals and $\ket{0}\!\bra{0}$ is diagonal, we can consider only the effect of $\mathcal{E}$ on diagonal elements, and say that

\begin{equation}
\begin{split}
\Tr\left[\ket{0}\!\bra{0}\mathcal{E}(\rho)\right] = &\Tr\left[\ket{0}\!\bra{0}\mathcal{E}\left(\sum_{k=0}^\infty p_{k,k} \ket{k}\!\bra{k}\right)\right]\\ = &\sum_{k=0}^\infty p_{k,k}\Tr\left[\ket{0}\!\bra{0}\mathcal{E}(\ket{k}\!\bra{k})\right],
\end{split}
\end{equation}

\noindent where here $p_{k,k}$ is the $k$-th diagonal element of $\rho$. 

We want our ultimate bounds to be in terms of the average photon number of the states. To relate the above quantity to this, we claim that 

\begin{equation} \label{eq:app:averageGreaterThanAverage}
\sum_{k=0}^\infty p_{k,k}\Tr\left[\ket{0}\!\bra{0}\mathcal{E}(\ket{k}\!\bra{k})\right] \geq \Tr\left[\ket{0}\!\bra{0}\mathcal{E}\left(\ket{\smash{\left\langle \hat{n}\right\rangle_\rho}}\!\bra{\smash{\left\langle \hat{n}\right\rangle_\rho}}\right)\right].
\end{equation}

\noindent That is to say, the average of the probabilities of losing each of many different photon number states is greater than the probability of losing one state of the average photon number (which we assume without loss of generality to be an integer).

Since a state is mapped to $\ket{0}\!\bra{0}$ if and only if it loses all of its photons, we can say that

\begin{equation} \label{eq:app:trEq1minusEtak}
\Tr\left[\ket{0}\!\bra{0}\mathcal{E}(\ket{k}\!\bra{k})\right] = \left(1-\eta\right)^k,
\end{equation}

We will consider first the case of $\rho$ being a mixture of only two Fock states, with weightings $p$ and $1-p$ and respective photon numbers $n$ and $m$. By using Eq. \ref{eq:app:trEq1minusEtak}, Eq. \ref{eq:app:averageGreaterThanAverage} then becomes

\begin{equation}
p(1-\eta)^n + (1-p)(1-\eta)^m \geq (1-\eta)^{p n + (1-p) m}.
\end{equation}

If we let $n = m + \delta$, then this simplifies to 

\begin{equation}
-y^p + p y - p + 1 \geq 0,
\end{equation}

\noindent where we have used $y \equiv (1-\eta)^\delta$. The claim then reduces to proving that this polynomial is satisfied for all $y, p \in [0,1]$.

Let $f(p) = -y^p + p y - p + 1$. We have $f(0) = f(1) = 0$. This function has a unique stationary point between 0 and 1, and the curvature $= -\left[\log(y)\right]^2 y^p$ is everywhere negative. Therefore $f(p) \geq 0$. This proves the claim for a bimodal initial state. The general claim follows by induction. 

We now have that 

\begin{equation} \label{eq:app:penultimate}
\Tr\left[\hat{E}_\checkmark\mathcal{E}(\rho)\right] \geq (1-\eta)^{\left\langle\hat{n}\right\rangle_\rho} = \left(1-\frac{\mu}{\left\langle\hat{n}\right\rangle_\rho}\right)^{\left\langle\hat{n}\right\rangle_\rho}.
\end{equation}

Since the average number of \emph{input} photon number is generally of the scale of dozens of orders of magnitude above unity, we may confidently take the limit of $\left\langle\hat{n}\right\rangle_\rho \rightarrow \infty$, which reduces Eq. \ref{eq:app:penultimate} to Eq. \ref{eq:EtickTrace}. 

\bibliographystyle{apsrev4-1}

\begin{thebibliography}{50}%
	\makeatletter
	\providecommand \@ifxundefined [1]{%
		\@ifx{#1\undefined}
	}%
	\providecommand \@ifnum [1]{%
		\ifnum #1\expandafter \@firstoftwo
		\else \expandafter \@secondoftwo
		\fi
	}%
	\providecommand \@ifx [1]{%
		\ifx #1\expandafter \@firstoftwo
		\else \expandafter \@secondoftwo
		\fi
	}%
	\providecommand \natexlab [1]{#1}%
	\providecommand \enquote  [1]{``#1''}%
	\providecommand \bibnamefont  [1]{#1}%
	\providecommand \bibfnamefont [1]{#1}%
	\providecommand \citenamefont [1]{#1}%
	\providecommand \href@noop [0]{\@secondoftwo}%
	\providecommand \href [0]{\begingroup \@sanitize@url \@href}%
	\providecommand \@href[1]{\@@startlink{#1}\@@href}%
	\providecommand \@@href[1]{\endgroup#1\@@endlink}%
	\providecommand \@sanitize@url [0]{\catcode `\\12\catcode `\$12\catcode
		`\&12\catcode `\#12\catcode `\^12\catcode `\_12\catcode `\%12\relax}%
	\providecommand \@@startlink[1]{}%
	\providecommand \@@endlink[0]{}%
	\providecommand \url  [0]{\begingroup\@sanitize@url \@url }%
	\providecommand \@url [1]{\endgroup\@href {#1}{\urlprefix }}%
	\providecommand \urlprefix  [0]{URL }%
	\providecommand \Eprint [0]{\href }%
	\providecommand \doibase [0]{http://dx.doi.org/}%
	\providecommand \selectlanguage [0]{\@gobble}%
	\providecommand \bibinfo  [0]{\@secondoftwo}%
	\providecommand \bibfield  [0]{\@secondoftwo}%
	\providecommand \translation [1]{[#1]}%
	\providecommand \BibitemOpen [0]{}%
	\providecommand \bibitemStop [0]{}%
	\providecommand \bibitemNoStop [0]{.\EOS\space}%
	\providecommand \EOS [0]{\spacefactor3000\relax}%
	\providecommand \BibitemShut  [1]{\csname bibitem#1\endcsname}%
	\let\auto@bib@innerbib\@empty
	%</preamble>
	\bibitem [{\citenamefont {Scarani}\ \emph {et~al.}(2009)\citenamefont
		{Scarani}, \citenamefont {Bechmann-Pasquinucci}, \citenamefont {Cerf},
		\citenamefont {Du??ek}, \citenamefont {L{\"{u}}tkenhaus},\ and\ \citenamefont
		{Peev}}]{Scarani2009TheDistribution}%
	\BibitemOpen
	\bibfield  {author} {\bibinfo {author} {\bibfnamefont {V.}~\bibnamefont
			{Scarani}}, \bibinfo {author} {\bibfnamefont {H.}~\bibnamefont
			{Bechmann-Pasquinucci}}, \bibinfo {author} {\bibfnamefont {N.~J.}\
			\bibnamefont {Cerf}}, \bibinfo {author} {\bibfnamefont {M.}~\bibnamefont
			{Du??ek}}, \bibinfo {author} {\bibfnamefont {N.}~\bibnamefont
			{L{\"{u}}tkenhaus}}, \ and\ \bibinfo {author} {\bibfnamefont
			{M.}~\bibnamefont {Peev}},\ }\href {\doibase 10.1103/RevModPhys.81.1301}
	{\bibfield  {journal} {\bibinfo  {journal} {Reviews of Modern Physics}\
		}\textbf {\bibinfo {volume} {81}},\ \bibinfo {pages} {1301} (\bibinfo {year}
		{2009})}\BibitemShut {NoStop}%
	\bibitem [{\citenamefont {Shor}(1997)}]{Shor1997Polynomial-TimeComputer}%
	\BibitemOpen
	\bibfield  {author} {\bibinfo {author} {\bibfnamefont {P.~W.}\ \bibnamefont
			{Shor}},\ }\href {\doibase 10.1137/S0097539795293172} {\bibfield  {journal}
		{\bibinfo  {journal} {SIAM Journal on Computing}\ }\textbf {\bibinfo {volume}
			{26}},\ \bibinfo {pages} {1484} (\bibinfo {year} {1997})}\BibitemShut
	{NoStop}%
	\bibitem [{\citenamefont {Renner}(2008)}]{renner_replaced}%
	\BibitemOpen
	\bibfield  {author} {\bibinfo {author} {\bibfnamefont {R.}~\bibnamefont
			{Renner}},\ }\href@noop {} {\bibfield  {journal} {\bibinfo  {journal}
			{International Journal of Quantum Information}\ }\textbf {\bibinfo {volume}
			{6}},\ \bibinfo {pages} {1} (\bibinfo {year} {2008})}\BibitemShut {NoStop}%
	\bibitem [{\citenamefont {Biham}\ and\ \citenamefont
		{Mor}(1997)}]{biham1997bounds}%
	\BibitemOpen
	\bibfield  {author} {\bibinfo {author} {\bibfnamefont {E.}~\bibnamefont
			{Biham}}\ and\ \bibinfo {author} {\bibfnamefont {T.}~\bibnamefont {Mor}},\
	}\href@noop {} {\bibfield  {journal} {\bibinfo  {journal} {Physical Review
				Letters}\ }\textbf {\bibinfo {volume} {79}},\ \bibinfo {pages} {4034}
		(\bibinfo {year} {1997})}\BibitemShut {NoStop}%
	\bibitem [{\citenamefont {Inamori}\ \emph {et~al.}(2007)\citenamefont
		{Inamori}, \citenamefont {L{\"{u}}tkenhaus},\ and\ \citenamefont
		{Mayers}}]{Inamori2007UnconditionalDistribution}%
	\BibitemOpen
	\bibfield  {author} {\bibinfo {author} {\bibfnamefont {H.}~\bibnamefont
			{Inamori}}, \bibinfo {author} {\bibfnamefont {N.}~\bibnamefont
			{L{\"{u}}tkenhaus}}, \ and\ \bibinfo {author} {\bibfnamefont
			{D.}~\bibnamefont {Mayers}},\ }\href {\doibase 10.1140/epjd/e2007-00010-4}
	{\bibfield  {journal} {\bibinfo  {journal} {European Physical Journal D}\
		}\textbf {\bibinfo {volume} {41}},\ \bibinfo {pages} {599} (\bibinfo {year}
		{2007})}\BibitemShut {NoStop}%
	\bibitem [{\citenamefont {Bennett}\ and\ \citenamefont
		{Brassard}(1984)}]{Bennett1984QuantumTossing.}%
	\BibitemOpen
	\bibfield  {author} {\bibinfo {author} {\bibfnamefont {C.~H.}\ \bibnamefont
			{Bennett}}\ and\ \bibinfo {author} {\bibfnamefont {G.}~\bibnamefont
			{Brassard}},\ }\href@noop {} {\bibfield  {journal} {\bibinfo  {journal}
			{International Conference on Computers, Systems {\&} Signal Processing}\
		}\textbf {\bibinfo {volume} {1}},\ \bibinfo {pages} {175} (\bibinfo {year}
		{1984})}\BibitemShut {NoStop}%
	\bibitem [{\citenamefont {Newton}\ \emph {et~al.}(2015)\citenamefont {Newton},
		\citenamefont {Everitt}, \citenamefont {Wilson},\ and\ \citenamefont
		{Varcoe}}]{Newton2015NovelDistribution}%
	\BibitemOpen
	\bibfield  {author} {\bibinfo {author} {\bibfnamefont {E.}~\bibnamefont
			{Newton}}, \bibinfo {author} {\bibfnamefont {M.~C.~J.}\ \bibnamefont
			{Everitt}}, \bibinfo {author} {\bibfnamefont {F.~L.}\ \bibnamefont {Wilson}},
		\ and\ \bibinfo {author} {\bibfnamefont {B.~T.~H.}\ \bibnamefont {Varcoe}},\
	}\href@noop {} {\  (\bibinfo {year} {2015})}\BibitemShut {NoStop}%
	\bibitem [{\citenamefont {Lydersen}\ \emph {et~al.}(2010)\citenamefont
		{Lydersen}, \citenamefont {Wiechers}, \citenamefont {Wittmann}, \citenamefont
		{Elser}, \citenamefont {Skaar},\ and\ \citenamefont
		{Makarov}}]{lydersen2010hacking}%
	\BibitemOpen
	\bibfield  {author} {\bibinfo {author} {\bibfnamefont {L.}~\bibnamefont
			{Lydersen}}, \bibinfo {author} {\bibfnamefont {C.}~\bibnamefont {Wiechers}},
		\bibinfo {author} {\bibfnamefont {C.}~\bibnamefont {Wittmann}}, \bibinfo
		{author} {\bibfnamefont {D.}~\bibnamefont {Elser}}, \bibinfo {author}
		{\bibfnamefont {J.}~\bibnamefont {Skaar}}, \ and\ \bibinfo {author}
		{\bibfnamefont {V.}~\bibnamefont {Makarov}},\ }\href@noop {} {\bibfield
		{journal} {\bibinfo  {journal} {Nature photonics}\ }\textbf {\bibinfo
			{volume} {4}},\ \bibinfo {pages} {686} (\bibinfo {year} {2010})}\BibitemShut
	{NoStop}%
	\bibitem [{\citenamefont {Comandar}\ \emph {et~al.}(2016)\citenamefont
		{Comandar}, \citenamefont {Lucamarini}, \citenamefont {Fr{\"{o}}hlich},
		\citenamefont {Dynes}, \citenamefont {Sharpe}, \citenamefont {Tam},
		\citenamefont {Yuan}, \citenamefont {Penty},\ and\ \citenamefont
		{Shields}}]{Comandar2016QuantumLasers}%
	\BibitemOpen
	\bibfield  {author} {\bibinfo {author} {\bibfnamefont {L.~C.}\ \bibnamefont
			{Comandar}}, \bibinfo {author} {\bibfnamefont {M.}~\bibnamefont
			{Lucamarini}}, \bibinfo {author} {\bibfnamefont {B.}~\bibnamefont
			{Fr{\"{o}}hlich}}, \bibinfo {author} {\bibfnamefont {J.~F.}\ \bibnamefont
			{Dynes}}, \bibinfo {author} {\bibfnamefont {A.~W.}\ \bibnamefont {Sharpe}},
		\bibinfo {author} {\bibfnamefont {S.~W.}\ \bibnamefont {Tam}}, \bibinfo
		{author} {\bibfnamefont {Z.~L.}\ \bibnamefont {Yuan}}, \bibinfo {author}
		{\bibfnamefont {R.~V.}\ \bibnamefont {Penty}}, \ and\ \bibinfo {author}
		{\bibfnamefont {A.~J.}\ \bibnamefont {Shields}},\ }\href {\doibase
		10.1038/nphoton.2016.50} {\bibfield  {journal} {\bibinfo  {journal} {Nature
				Photonics}\ }\textbf {\bibinfo {volume} {10}},\ \bibinfo {pages} {312}
		(\bibinfo {year} {2016})}\BibitemShut {NoStop}%
	\bibitem [{\citenamefont {Scarani}\ \emph {et~al.}(2004)\citenamefont
		{Scarani}, \citenamefont {Acin}, \citenamefont {Ribordy},\ and\ \citenamefont
		{Gisin}}]{scarani2004quantum_replace}%
	\BibitemOpen
	\bibfield  {author} {\bibinfo {author} {\bibfnamefont {V.}~\bibnamefont
			{Scarani}}, \bibinfo {author} {\bibfnamefont {A.}~\bibnamefont {Acin}},
		\bibinfo {author} {\bibfnamefont {G.}~\bibnamefont {Ribordy}}, \ and\
		\bibinfo {author} {\bibfnamefont {N.}~\bibnamefont {Gisin}},\ }\href@noop {}
	{\bibfield  {journal} {\bibinfo  {journal} {Physical review letters}\
		}\textbf {\bibinfo {volume} {92}},\ \bibinfo {pages} {057901} (\bibinfo
		{year} {2004})}\BibitemShut {NoStop}%
	\bibitem [{\citenamefont {Curty}\ \emph {et~al.}(2014)\citenamefont {Curty},
		\citenamefont {Xu}, \citenamefont {Cui}, \citenamefont {Lim}, \citenamefont
		{Tamaki},\ and\ \citenamefont {Lo}}]{Curty2014Finite-keyDistribution}%
	\BibitemOpen
	\bibfield  {author} {\bibinfo {author} {\bibfnamefont {M.}~\bibnamefont
			{Curty}}, \bibinfo {author} {\bibfnamefont {F.}~\bibnamefont {Xu}}, \bibinfo
		{author} {\bibfnamefont {W.}~\bibnamefont {Cui}}, \bibinfo {author}
		{\bibfnamefont {C.~C.~W.}\ \bibnamefont {Lim}}, \bibinfo {author}
		{\bibfnamefont {K.}~\bibnamefont {Tamaki}}, \ and\ \bibinfo {author}
		{\bibfnamefont {H.~K.}\ \bibnamefont {Lo}},\ }\href {\doibase
		10.1038/ncomms4732} {\bibfield  {journal} {\bibinfo  {journal} {Nature
				Communications}\ }\textbf {\bibinfo {volume} {5}},\ \bibinfo {pages} {1}
		(\bibinfo {year} {2014})}\BibitemShut {NoStop}%
	\bibitem [{\citenamefont {Braunstein}\ and\ \citenamefont
		{Pirandola}(2012)}]{Braunstein2012Side-channel-freeDistribution}%
	\BibitemOpen
	\bibfield  {author} {\bibinfo {author} {\bibfnamefont {S.~L.}\ \bibnamefont
			{Braunstein}}\ and\ \bibinfo {author} {\bibfnamefont {S.}~\bibnamefont
			{Pirandola}},\ }\href {\doibase 10.1103/PhysRevLett.108.130502} {\bibfield
		{journal} {\bibinfo  {journal} {Physical Review Letters}\ }\textbf {\bibinfo
			{volume} {108}},\ \bibinfo {pages} {1} (\bibinfo {year} {2012})}\BibitemShut
	{NoStop}%
	\bibitem [{\citenamefont {Gerhardt}\ \emph {et~al.}(2011)\citenamefont
		{Gerhardt}, \citenamefont {Liu}, \citenamefont {Lamas-Linares}, \citenamefont
		{Skaar}, \citenamefont {Kurtsiefer},\ and\ \citenamefont
		{Makarov}}]{Gerhardt2011Full-fieldSystem}%
	\BibitemOpen
	\bibfield  {author} {\bibinfo {author} {\bibfnamefont {I.}~\bibnamefont
			{Gerhardt}}, \bibinfo {author} {\bibfnamefont {Q.}~\bibnamefont {Liu}},
		\bibinfo {author} {\bibfnamefont {A.~A.}\ \bibnamefont {Lamas-Linares}},
		\bibinfo {author} {\bibfnamefont {J.}~\bibnamefont {Skaar}}, \bibinfo
		{author} {\bibfnamefont {C.}~\bibnamefont {Kurtsiefer}}, \ and\ \bibinfo
		{author} {\bibfnamefont {V.}~\bibnamefont {Makarov}},\ }\href {\doibase
		10.1038/ncomms1348} {\bibfield  {journal} {\bibinfo  {journal} {Nature
				Communications}\ }\textbf {\bibinfo {volume} {2}} (\bibinfo {year} {2011}),\
		10.1038/ncomms1348}\BibitemShut {NoStop}%
	\bibitem [{\citenamefont {Xu}\ \emph {et~al.}(2010)\citenamefont {Xu},
		\citenamefont {Qi},\ and\ \citenamefont {Lo}}]{Xu2010ExperimentalSystem}%
	\BibitemOpen
	\bibfield  {author} {\bibinfo {author} {\bibfnamefont {F.}~\bibnamefont
			{Xu}}, \bibinfo {author} {\bibfnamefont {B.}~\bibnamefont {Qi}}, \ and\
		\bibinfo {author} {\bibfnamefont {H.~K.}\ \bibnamefont {Lo}},\ }\href
	{\doibase 10.1088/1367-2630/12/11/113026} {\bibfield  {journal} {\bibinfo
			{journal} {New Journal of Physics}\ }\textbf {\bibinfo {volume} {12}}
		(\bibinfo {year} {2010}),\ 10.1088/1367-2630/12/11/113026}\BibitemShut
	{NoStop}%
	\bibitem [{\citenamefont {Zhao}\ \emph {et~al.}(2008)\citenamefont {Zhao},
		\citenamefont {Fung}, \citenamefont {Qi}, \citenamefont {Chen},\ and\
		\citenamefont {Lo}}]{Zhao2008QuantumSystems}%
	\BibitemOpen
	\bibfield  {author} {\bibinfo {author} {\bibfnamefont {Y.}~\bibnamefont
			{Zhao}}, \bibinfo {author} {\bibfnamefont {C.~H.~F.}\ \bibnamefont {Fung}},
		\bibinfo {author} {\bibfnamefont {B.}~\bibnamefont {Qi}}, \bibinfo {author}
		{\bibfnamefont {C.}~\bibnamefont {Chen}}, \ and\ \bibinfo {author}
		{\bibfnamefont {H.~K.}\ \bibnamefont {Lo}},\ }\href {\doibase
		10.1103/PhysRevA.78.042333} {\bibfield  {journal} {\bibinfo  {journal}
			{Physical Review A - Atomic, Molecular, and Optical Physics}\ }\textbf
		{\bibinfo {volume} {78}},\ \bibinfo {pages} {1} (\bibinfo {year}
		{2008})}\BibitemShut {NoStop}%
	\bibitem [{\citenamefont {Wang}(2005)}]{Wang2005BeatingCryptography}%
	\BibitemOpen
	\bibfield  {author} {\bibinfo {author} {\bibfnamefont {X.~B.}\ \bibnamefont
			{Wang}},\ }\href {\doibase 10.1103/PhysRevLett.94.230503} {\bibfield
		{journal} {\bibinfo  {journal} {Physical Review Letters}\ }\textbf {\bibinfo
			{volume} {94}},\ \bibinfo {pages} {1} (\bibinfo {year} {2005})}\BibitemShut
	{NoStop}%
	\bibitem [{\citenamefont {Gisin}\ \emph {et~al.}(2006)\citenamefont {Gisin},
		\citenamefont {Fasel}, \citenamefont {Kraus}, \citenamefont {Zbinden},\ and\
		\citenamefont {Ribordy}}]{gisin2006trojan}%
	\BibitemOpen
	\bibfield  {author} {\bibinfo {author} {\bibfnamefont {N.}~\bibnamefont
			{Gisin}}, \bibinfo {author} {\bibfnamefont {S.}~\bibnamefont {Fasel}},
		\bibinfo {author} {\bibfnamefont {B.}~\bibnamefont {Kraus}}, \bibinfo
		{author} {\bibfnamefont {H.}~\bibnamefont {Zbinden}}, \ and\ \bibinfo
		{author} {\bibfnamefont {G.}~\bibnamefont {Ribordy}},\ }\href@noop {}
	{\bibfield  {journal} {\bibinfo  {journal} {Physical Review A}\ }\textbf
		{\bibinfo {volume} {73}},\ \bibinfo {pages} {022320} (\bibinfo {year}
		{2006})}\BibitemShut {NoStop}%
	\bibitem [{\citenamefont {Deng}\ \emph {et~al.}(2005)\citenamefont {Deng},
		\citenamefont {Li}, \citenamefont {Zhou},\ and\ \citenamefont
		{Zhang}}]{deng2005improving}%
	\BibitemOpen
	\bibfield  {author} {\bibinfo {author} {\bibfnamefont {F.-G.}\ \bibnamefont
			{Deng}}, \bibinfo {author} {\bibfnamefont {X.-H.}\ \bibnamefont {Li}},
		\bibinfo {author} {\bibfnamefont {H.-Y.}\ \bibnamefont {Zhou}}, \ and\
		\bibinfo {author} {\bibfnamefont {Z.-j.}\ \bibnamefont {Zhang}},\ }\href@noop
	{} {\bibfield  {journal} {\bibinfo  {journal} {Physical Review A}\ }\textbf
		{\bibinfo {volume} {72}},\ \bibinfo {pages} {044302} (\bibinfo {year}
		{2005})}\BibitemShut {NoStop}%
	\bibitem [{\citenamefont {Lucamarini}\ \emph {et~al.}(2015)\citenamefont
		{Lucamarini}, \citenamefont {Choi}, \citenamefont {Ward}, \citenamefont
		{Dynes}, \citenamefont {Yuan},\ and\ \citenamefont
		{Shields}}]{Lucamarini2015PracticalDistribution}%
	\BibitemOpen
	\bibfield  {author} {\bibinfo {author} {\bibfnamefont {M.}~\bibnamefont
			{Lucamarini}}, \bibinfo {author} {\bibfnamefont {I.}~\bibnamefont {Choi}},
		\bibinfo {author} {\bibfnamefont {M.~B.}\ \bibnamefont {Ward}}, \bibinfo
		{author} {\bibfnamefont {J.~F.}\ \bibnamefont {Dynes}}, \bibinfo {author}
		{\bibfnamefont {Z.~L.}\ \bibnamefont {Yuan}}, \ and\ \bibinfo {author}
		{\bibfnamefont {A.~J.}\ \bibnamefont {Shields}},\ }\href {\doibase
		10.1103/PhysRevX.5.031030} {\bibfield  {journal} {\bibinfo  {journal}
			{Physical Review X}\ }\textbf {\bibinfo {volume} {5}},\ \bibinfo {pages} {1}
		(\bibinfo {year} {2015})}\BibitemShut {NoStop}%
	\bibitem [{\citenamefont {Jain}\ \emph {et~al.}(2015)\citenamefont {Jain},
		\citenamefont {Stiller}, \citenamefont {Khan}, \citenamefont {Makarov},
		\citenamefont {Marquardt},\ and\ \citenamefont {Leuchs}}]{jain2015risk}%
	\BibitemOpen
	\bibfield  {author} {\bibinfo {author} {\bibfnamefont {N.}~\bibnamefont
			{Jain}}, \bibinfo {author} {\bibfnamefont {B.}~\bibnamefont {Stiller}},
		\bibinfo {author} {\bibfnamefont {I.}~\bibnamefont {Khan}}, \bibinfo {author}
		{\bibfnamefont {V.}~\bibnamefont {Makarov}}, \bibinfo {author} {\bibfnamefont
			{C.}~\bibnamefont {Marquardt}}, \ and\ \bibinfo {author} {\bibfnamefont
			{G.}~\bibnamefont {Leuchs}},\ }\href@noop {} {\bibfield  {journal} {\bibinfo
			{journal} {IEEE Journal of Selected Topics in Quantum Electronics}\ }\textbf
		{\bibinfo {volume} {21}},\ \bibinfo {pages} {168} (\bibinfo {year}
		{2015})}\BibitemShut {NoStop}%
	\bibitem [{\citenamefont {Jain}\ \emph {et~al.}(2014)\citenamefont {Jain},
		\citenamefont {Anisimova}, \citenamefont {Khan}, \citenamefont {Makarov},
		\citenamefont {Marquardt},\ and\ \citenamefont {Leuchs}}]{jain2014trojan}%
	\BibitemOpen
	\bibfield  {author} {\bibinfo {author} {\bibfnamefont {N.}~\bibnamefont
			{Jain}}, \bibinfo {author} {\bibfnamefont {E.}~\bibnamefont {Anisimova}},
		\bibinfo {author} {\bibfnamefont {I.}~\bibnamefont {Khan}}, \bibinfo {author}
		{\bibfnamefont {V.}~\bibnamefont {Makarov}}, \bibinfo {author} {\bibfnamefont
			{C.}~\bibnamefont {Marquardt}}, \ and\ \bibinfo {author} {\bibfnamefont
			{G.}~\bibnamefont {Leuchs}},\ }\href@noop {} {\bibfield  {journal} {\bibinfo
			{journal} {New Journal of Physics}\ }\textbf {\bibinfo {volume} {16}},\
		\bibinfo {pages} {123030} (\bibinfo {year} {2014})}\BibitemShut {NoStop}%
	\bibitem [{\citenamefont {Sajeed}\ \emph {et~al.}(2015)\citenamefont {Sajeed},
		\citenamefont {Radchenko}, \citenamefont {Kaiser}, \citenamefont {Bourgoin},
		\citenamefont {Pappa}, \citenamefont {Monat}, \citenamefont {Legr{\'{e}}},\
		and\ \citenamefont {Makarov}}]{Sajeed2015AttacksTossing}%
	\BibitemOpen
	\bibfield  {author} {\bibinfo {author} {\bibfnamefont {S.}~\bibnamefont
			{Sajeed}}, \bibinfo {author} {\bibfnamefont {I.}~\bibnamefont {Radchenko}},
		\bibinfo {author} {\bibfnamefont {S.}~\bibnamefont {Kaiser}}, \bibinfo
		{author} {\bibfnamefont {J.~P.}\ \bibnamefont {Bourgoin}}, \bibinfo {author}
		{\bibfnamefont {A.}~\bibnamefont {Pappa}}, \bibinfo {author} {\bibfnamefont
			{L.}~\bibnamefont {Monat}}, \bibinfo {author} {\bibfnamefont
			{M.}~\bibnamefont {Legr{\'{e}}}}, \ and\ \bibinfo {author} {\bibfnamefont
			{V.}~\bibnamefont {Makarov}},\ }\href {\doibase 10.1103/PhysRevA.91.032326}
	{\bibfield  {journal} {\bibinfo  {journal} {Physical Review A - Atomic,
				Molecular, and Optical Physics}\ }\textbf {\bibinfo {volume} {91}},\ \bibinfo
		{pages} {1} (\bibinfo {year} {2015})}\BibitemShut {NoStop}%
	\bibitem [{\citenamefont {Gottesman}\ \emph {et~al.}(2004)\citenamefont
		{Gottesman}, \citenamefont {Lo}, \citenamefont {Lutkenhaus},\ and\
		\citenamefont {Preskill}}]{gottesman_replaced}%
	\BibitemOpen
	\bibfield  {author} {\bibinfo {author} {\bibfnamefont {D.}~\bibnamefont
			{Gottesman}}, \bibinfo {author} {\bibfnamefont {H.-K.}\ \bibnamefont {Lo}},
		\bibinfo {author} {\bibfnamefont {N.}~\bibnamefont {Lutkenhaus}}, \ and\
		\bibinfo {author} {\bibfnamefont {J.}~\bibnamefont {Preskill}},\ }in\
	\href@noop {} {\emph {\bibinfo {booktitle} {Information Theory, 2004. ISIT
				2004. Proceedings. International Symposium on}}}\ (\bibinfo {organization}
	{IEEE},\ \bibinfo {year} {2004})\ p.\ \bibinfo {pages} {136}\BibitemShut
	{NoStop}%
	\bibitem [{\citenamefont {Driscoll}\ \emph {et~al.}(1991)\citenamefont
		{Driscoll}, \citenamefont {Calo},\ and\ \citenamefont
		{Lawandy}}]{Driscoll1991ExplainingFuse.}%
	\BibitemOpen
	\bibfield  {author} {\bibinfo {author} {\bibfnamefont {T.~J.}\ \bibnamefont
			{Driscoll}}, \bibinfo {author} {\bibfnamefont {J.~M.}\ \bibnamefont {Calo}},
		\ and\ \bibinfo {author} {\bibfnamefont {N.~M.}\ \bibnamefont {Lawandy}},\
	}\href {\doibase 10.1364/OL.16.001046} {\bibfield  {journal} {\bibinfo
			{journal} {Optics letters}\ }\textbf {\bibinfo {volume} {16}},\ \bibinfo
		{pages} {1046} (\bibinfo {year} {1991})}\BibitemShut {NoStop}%
	\bibitem [{\citenamefont {Carr}\ \emph {et~al.}(2003)\citenamefont {Carr},
		\citenamefont {Radousky},\ and\ \citenamefont
		{Demos}}]{Carr2003WavelengthMechanisms}%
	\BibitemOpen
	\bibfield  {author} {\bibinfo {author} {\bibfnamefont {C.~W.}\ \bibnamefont
			{Carr}}, \bibinfo {author} {\bibfnamefont {H.~B.}\ \bibnamefont {Radousky}},
		\ and\ \bibinfo {author} {\bibfnamefont {S.~G.}\ \bibnamefont {Demos}},\
	}\href {\doibase 10.1103/PhysRevLett.91.127402} {\bibfield  {journal}
		{\bibinfo  {journal} {Physical Review Letters}\ }\textbf {\bibinfo {volume}
			{91}},\ \bibinfo {pages} {127402} (\bibinfo {year} {2003})}\BibitemShut
	{NoStop}%
	\bibitem [{\citenamefont {Lloyd}(2008)}]{Lloyd2008EnhancedIllumination}%
	\BibitemOpen
	\bibfield  {author} {\bibinfo {author} {\bibfnamefont {S.}~\bibnamefont
			{Lloyd}},\ }\href {\doibase 10.1126/science.1160627} {\bibfield  {journal}
		{\bibinfo  {journal} {Science}\ }\textbf {\bibinfo {volume} {321}},\ \bibinfo
		{pages} {1463} (\bibinfo {year} {2008})}\BibitemShut {NoStop}%
	\bibitem [{\citenamefont {Nielsen}\ and\ \citenamefont
		{Chuang}(2002)}]{nielsen2002quantum}%
	\BibitemOpen
	\bibfield  {author} {\bibinfo {author} {\bibfnamefont {M.~A.}\ \bibnamefont
			{Nielsen}}\ and\ \bibinfo {author} {\bibfnamefont {I.}~\bibnamefont
			{Chuang}},\ }\href@noop {} {\emph {\bibinfo {title} {Quantum computation and
				quantum information}}}\ (\bibinfo  {publisher} {AAPT},\ \bibinfo {year}
	{2002})\BibitemShut {NoStop}%
	\bibitem [{\citenamefont {Braunstein}(2005)}]{Braunstein2005SqueezingResource}%
	\BibitemOpen
	\bibfield  {author} {\bibinfo {author} {\bibfnamefont {S.~L.}\ \bibnamefont
			{Braunstein}},\ }\href {\doibase 10.1103/PhysRevA.71.055801} {\bibfield
		{journal} {\bibinfo  {journal} {Physical Review A - Atomic, Molecular, and
				Optical Physics}\ }\textbf {\bibinfo {volume} {71}},\ \bibinfo {pages} {8}
		(\bibinfo {year} {2005})}\BibitemShut {NoStop}%
	\bibitem [{\citenamefont {Lasota}\ \emph {et~al.}(2017)\citenamefont {Lasota},
		\citenamefont {Filip},\ and\ \citenamefont
		{Usenko}}]{Lasota2017RobustnessNoise}%
	\BibitemOpen
	\bibfield  {author} {\bibinfo {author} {\bibfnamefont {M.}~\bibnamefont
			{Lasota}}, \bibinfo {author} {\bibfnamefont {R.}~\bibnamefont {Filip}}, \
		and\ \bibinfo {author} {\bibfnamefont {V.~C.}\ \bibnamefont {Usenko}},\
	}\href {\doibase 10.1103/PhysRevA.95.062312} {\ \textbf {\bibinfo {volume}
			{062312}},\ \bibinfo {pages} {1} (\bibinfo {year} {2017})}\BibitemShut
	{NoStop}%
	\bibitem [{\citenamefont {Glauber}(1963)}]{Glauber1963CoherentField}%
	\BibitemOpen
	\bibfield  {author} {\bibinfo {author} {\bibfnamefont {R.~J.}\ \bibnamefont
			{Glauber}},\ }\href@noop {} {\bibfield  {journal} {\bibinfo  {journal}
			{Physical Review}\ }\textbf {\bibinfo {volume} {131}},\ \bibinfo {pages}
		{2766} (\bibinfo {year} {1963})}\BibitemShut {NoStop}%
	\bibitem [{\citenamefont {Koashi}(2009)}]{Koashi2009SimpleComplementarity}%
	\BibitemOpen
	\bibfield  {author} {\bibinfo {author} {\bibfnamefont {M.}~\bibnamefont
			{Koashi}},\ }\href {\doibase 10.1088/1367-2630/11/4/045018} {\bibfield
		{journal} {\bibinfo  {journal} {New Journal of Physics}\ }\textbf {\bibinfo
			{volume} {11}} (\bibinfo {year} {2009}),\
		10.1088/1367-2630/11/4/045018}\BibitemShut {NoStop}%
	\bibitem [{\citenamefont {Tamaki}\ \emph {et~al.}(2003)\citenamefont {Tamaki},
		\citenamefont {Koashi},\ and\ \citenamefont
		{Imoto}}]{Tamaki2003UnconditionallyStates}%
	\BibitemOpen
	\bibfield  {author} {\bibinfo {author} {\bibfnamefont {K.}~\bibnamefont
			{Tamaki}}, \bibinfo {author} {\bibfnamefont {M.}~\bibnamefont {Koashi}}, \
		and\ \bibinfo {author} {\bibfnamefont {N.}~\bibnamefont {Imoto}},\ }\href
	{\doibase 10.1103/PhysRevLett.90.167904} {\bibfield  {journal} {\bibinfo
			{journal} {Physical Review Letters}\ }\textbf {\bibinfo {volume} {90}},\
		\bibinfo {pages} {167904} (\bibinfo {year} {2003})}\BibitemShut {NoStop}%
	\bibitem [{Note1()}]{Note1}%
	\BibitemOpen
	\bibinfo {note} {This may be seen by considering Ref. \cite
		{gottesman_replaced}, section VIII. The purified states corresponding to each
		basis are as defined in Ref. \cite {Lucamarini2015PracticalDistribution},
		Appendix B. From these it may be seen that $1-2\Delta $ is equal to the
		\protect \emph {average} fidelity between a state being emitted in the $X$
		basis and one in the $Y$ basis. By the symmetry and unitary invariance of the
		fidelity function, this reduces to finding the fidelity between only the
		states corresponding to $\theta = 0$ and $\theta = \pi /2$.}\BibitemShut
	{Stop}%
	\bibitem [{\citenamefont {Banchi}\ \emph {et~al.}(2015)\citenamefont {Banchi},
		\citenamefont {Braunstein},\ and\ \citenamefont
		{Pirandola}}]{Banchi2015QuantumStates}%
	\BibitemOpen
	\bibfield  {author} {\bibinfo {author} {\bibfnamefont {L.}~\bibnamefont
			{Banchi}}, \bibinfo {author} {\bibfnamefont {S.~L.}\ \bibnamefont
			{Braunstein}}, \ and\ \bibinfo {author} {\bibfnamefont {S.}~\bibnamefont
			{Pirandola}},\ }\href {\doibase 10.1103/PhysRevLett.115.260501} {\bibfield
		{journal} {\bibinfo  {journal} {Physical Review Letters}\ }\textbf {\bibinfo
			{volume} {115}},\ \bibinfo {pages} {1} (\bibinfo {year} {2015})}\BibitemShut
	{NoStop}%
	\bibitem [{\citenamefont {Lo}\ \emph {et~al.}(2005)\citenamefont {Lo},
		\citenamefont {Chau},\ and\ \citenamefont {Ardehali}}]{Lo_replaced}%
	\BibitemOpen
	\bibfield  {author} {\bibinfo {author} {\bibfnamefont {H.-K.}\ \bibnamefont
			{Lo}}, \bibinfo {author} {\bibfnamefont {H.~F.}\ \bibnamefont {Chau}}, \ and\
		\bibinfo {author} {\bibfnamefont {M.}~\bibnamefont {Ardehali}},\ }\href@noop
	{} {\bibfield  {journal} {\bibinfo  {journal} {Journal of Cryptology}\
		}\textbf {\bibinfo {volume} {18}},\ \bibinfo {pages} {133} (\bibinfo {year}
		{2005})}\BibitemShut {NoStop}%
	\bibitem [{\citenamefont {Simon}\ \emph {et~al.}(2007)\citenamefont {Simon},
		\citenamefont {De~Riedmatten}, \citenamefont {Afzelius}, \citenamefont
		{Sangouard}, \citenamefont {Zbinden},\ and\ \citenamefont
		{Gisin}}]{Simon2007QuantumMemories}%
	\BibitemOpen
	\bibfield  {author} {\bibinfo {author} {\bibfnamefont {C.}~\bibnamefont
			{Simon}}, \bibinfo {author} {\bibfnamefont {H.}~\bibnamefont
			{De~Riedmatten}}, \bibinfo {author} {\bibfnamefont {M.}~\bibnamefont
			{Afzelius}}, \bibinfo {author} {\bibfnamefont {N.}~\bibnamefont {Sangouard}},
		\bibinfo {author} {\bibfnamefont {H.}~\bibnamefont {Zbinden}}, \ and\
		\bibinfo {author} {\bibfnamefont {N.}~\bibnamefont {Gisin}},\ }\href
	{\doibase 10.1103/PhysRevLett.98.190503} {\bibfield  {journal} {\bibinfo
			{journal} {Physical Review Letters}\ }\textbf {\bibinfo {volume} {98}},\
		\bibinfo {pages} {1} (\bibinfo {year} {2007})}\BibitemShut {NoStop}%
	\bibitem [{\citenamefont {Nemoto}\ \emph {et~al.}(2016)\citenamefont {Nemoto},
		\citenamefont {Trupke}, \citenamefont {Devitt}, \citenamefont
		{Scharfenberger}, \citenamefont {Buczak}, \citenamefont {Schmiedmayer},\ and\
		\citenamefont {Munro}}]{nemoto_replaced}%
	\BibitemOpen
	\bibfield  {author} {\bibinfo {author} {\bibfnamefont {K.}~\bibnamefont
			{Nemoto}}, \bibinfo {author} {\bibfnamefont {M.}~\bibnamefont {Trupke}},
		\bibinfo {author} {\bibfnamefont {S.~J.}\ \bibnamefont {Devitt}}, \bibinfo
		{author} {\bibfnamefont {B.}~\bibnamefont {Scharfenberger}}, \bibinfo
		{author} {\bibfnamefont {K.}~\bibnamefont {Buczak}}, \bibinfo {author}
		{\bibfnamefont {J.}~\bibnamefont {Schmiedmayer}}, \ and\ \bibinfo {author}
		{\bibfnamefont {W.~J.}\ \bibnamefont {Munro}},\ }\href@noop {} {\bibfield
		{journal} {\bibinfo  {journal} {Scientific reports}\ }\textbf {\bibinfo
			{volume} {6}} (\bibinfo {year} {2016})}\BibitemShut {NoStop}%
	\bibitem [{\citenamefont {Zwerger}\ \emph {et~al.}(2012)\citenamefont
		{Zwerger}, \citenamefont {D{\"{u}}r},\ and\ \citenamefont
		{Briegel}}]{Zwerger2012Measurement-basedRepeaters}%
	\BibitemOpen
	\bibfield  {author} {\bibinfo {author} {\bibfnamefont {M.}~\bibnamefont
			{Zwerger}}, \bibinfo {author} {\bibfnamefont {W.}~\bibnamefont {D{\"{u}}r}},
		\ and\ \bibinfo {author} {\bibfnamefont {H.~J.}\ \bibnamefont {Briegel}},\
	}\href {\doibase 10.1103/PhysRevA.85.062326} {\bibfield  {journal} {\bibinfo
			{journal} {Physical Review A - Atomic, Molecular, and Optical Physics}\
		}\textbf {\bibinfo {volume} {85}},\ \bibinfo {pages} {1} (\bibinfo {year}
		{2012})}\BibitemShut {NoStop}%
	\bibitem [{\citenamefont {Vinay}\ and\ \citenamefont
		{Kok}(2017)}]{Vinay2017PracticalCommunication}%
	\BibitemOpen
	\bibfield  {author} {\bibinfo {author} {\bibfnamefont {S.~E.}\ \bibnamefont
			{Vinay}}\ and\ \bibinfo {author} {\bibfnamefont {P.}~\bibnamefont {Kok}},\
	}\href {\doibase 10.1103/PhysRevA.95.052336} {\bibfield  {journal} {\bibinfo
			{journal} {Physical Review A}\ }\textbf {\bibinfo {volume} {95}},\ \bibinfo
		{pages} {1} (\bibinfo {year} {2017})}\BibitemShut {NoStop}%
	\bibitem [{\citenamefont {Deutsch}\ \emph {et~al.}(1996)\citenamefont
		{Deutsch}, \citenamefont {Ekert}, \citenamefont {Jozsa}, \citenamefont
		{Macchiavello}, \citenamefont {Popescu},\ and\ \citenamefont
		{Sanpera}}]{Deutsch1996QuantumChannels}%
	\BibitemOpen
	\bibfield  {author} {\bibinfo {author} {\bibfnamefont {D.}~\bibnamefont
			{Deutsch}}, \bibinfo {author} {\bibfnamefont {A.}~\bibnamefont {Ekert}},
		\bibinfo {author} {\bibfnamefont {R.}~\bibnamefont {Jozsa}}, \bibinfo
		{author} {\bibfnamefont {C.}~\bibnamefont {Macchiavello}}, \bibinfo {author}
		{\bibfnamefont {S.}~\bibnamefont {Popescu}}, \ and\ \bibinfo {author}
		{\bibfnamefont {A.}~\bibnamefont {Sanpera}},\ }\href {\doibase
		10.1103/PhysRevLett.77.2818} {\bibfield  {journal} {\bibinfo  {journal}
			{Phys. Rev. Lett.}\ }\textbf {\bibinfo {volume} {77}},\ \bibinfo {pages}
		{2818} (\bibinfo {year} {1996})}\BibitemShut {NoStop}%
	\bibitem [{\citenamefont {Renner}\ and\ \citenamefont
		{K{\"o}nig}(2005)}]{renner2005universally}%
	\BibitemOpen
	\bibfield  {author} {\bibinfo {author} {\bibfnamefont {R.}~\bibnamefont
			{Renner}}\ and\ \bibinfo {author} {\bibfnamefont {R.}~\bibnamefont
			{K{\"o}nig}},\ }in\ \href@noop {} {\emph {\bibinfo {booktitle} {Theory of
				Cryptography Conference}}}\ (\bibinfo {organization} {Springer},\ \bibinfo
	{year} {2005})\ pp.\ \bibinfo {pages} {407--425}\BibitemShut {NoStop}%
	\bibitem [{\citenamefont {Duan}\ \emph {et~al.}(2001)\citenamefont {Duan},
		\citenamefont {Lukin}, \citenamefont {Cirac},\ and\ \citenamefont
		{Zoller}}]{Duan2001Long-distanceOptics.}%
	\BibitemOpen
	\bibfield  {author} {\bibinfo {author} {\bibfnamefont {L.~M.}\ \bibnamefont
			{Duan}}, \bibinfo {author} {\bibfnamefont {M.~D.}\ \bibnamefont {Lukin}},
		\bibinfo {author} {\bibfnamefont {J.~I.}\ \bibnamefont {Cirac}}, \ and\
		\bibinfo {author} {\bibfnamefont {P.}~\bibnamefont {Zoller}},\ }\href
	{\doibase 10.1038/35106500} {\bibfield  {journal} {\bibinfo  {journal}
			{Nature}\ }\textbf {\bibinfo {volume} {414}},\ \bibinfo {pages} {413}
		(\bibinfo {year} {2001})}\BibitemShut {NoStop}%
	\bibitem [{\citenamefont {Piparo}\ and\ \citenamefont
		{Razavi}(2015)}]{Piparo2015Long-DistanceDistribution}%
	\BibitemOpen
	\bibfield  {author} {\bibinfo {author} {\bibfnamefont {N.~L.}\ \bibnamefont
			{Piparo}}\ and\ \bibinfo {author} {\bibfnamefont {M.}~\bibnamefont
			{Razavi}},\ }\href {\doibase 10.1109/JSTQE.2014.2364129} {\bibfield
		{journal} {\bibinfo  {journal} {IEEE Journal on Selected Topics in Quantum
				Electronics}\ }\textbf {\bibinfo {volume} {21}} (\bibinfo {year} {2015}),\
		10.1109/JSTQE.2014.2364129}\BibitemShut {NoStop}%
	\bibitem [{\citenamefont {Epping}\ \emph {et~al.}(2016)\citenamefont {Epping},
		\citenamefont {Kampermann},\ and\ \citenamefont
		{Bru{\ss}}}]{epping_replaced}%
	\BibitemOpen
	\bibfield  {author} {\bibinfo {author} {\bibfnamefont {M.}~\bibnamefont
			{Epping}}, \bibinfo {author} {\bibfnamefont {H.}~\bibnamefont {Kampermann}},
		\ and\ \bibinfo {author} {\bibfnamefont {D.}~\bibnamefont {Bru{\ss}}},\
	}\href@noop {} {\bibfield  {journal} {\bibinfo  {journal} {New Journal of
				Physics}\ }\textbf {\bibinfo {volume} {18}},\ \bibinfo {pages} {103052}
		(\bibinfo {year} {2016})}\BibitemShut {NoStop}%
	\bibitem [{\citenamefont {Piparo}\ \emph {et~al.}(2017)\citenamefont {Piparo},
		\citenamefont {Razavi},\ and\ \citenamefont {Munro}}]{piparo_replaced}%
	\BibitemOpen
	\bibfield  {author} {\bibinfo {author} {\bibfnamefont {N.~L.}\ \bibnamefont
			{Piparo}}, \bibinfo {author} {\bibfnamefont {M.}~\bibnamefont {Razavi}}, \
		and\ \bibinfo {author} {\bibfnamefont {W.~J.}\ \bibnamefont {Munro}},\
	}\href@noop {} {\bibfield  {journal} {\bibinfo  {journal} {Physical Review
				A}\ }\textbf {\bibinfo {volume} {96}},\ \bibinfo {pages} {052313} (\bibinfo
		{year} {2017})}\BibitemShut {NoStop}%
	\bibitem [{\citenamefont {Kok}\ \emph {et~al.}(2007)\citenamefont {Kok},
		\citenamefont {Munro}, \citenamefont {Nemoto}, \citenamefont {Ralph},
		\citenamefont {Dowling},\ and\ \citenamefont
		{Milburn}}]{Kok2007LinearQubits}%
	\BibitemOpen
	\bibfield  {author} {\bibinfo {author} {\bibfnamefont {P.}~\bibnamefont
			{Kok}}, \bibinfo {author} {\bibfnamefont {W.~J.}\ \bibnamefont {Munro}},
		\bibinfo {author} {\bibfnamefont {K.}~\bibnamefont {Nemoto}}, \bibinfo
		{author} {\bibfnamefont {T.~C.}\ \bibnamefont {Ralph}}, \bibinfo {author}
		{\bibfnamefont {J.~P.}\ \bibnamefont {Dowling}}, \ and\ \bibinfo {author}
		{\bibfnamefont {G.~J.}\ \bibnamefont {Milburn}},\ }\href {\doibase
		10.1103/RevModPhys.79.135} {\bibfield  {journal} {\bibinfo  {journal}
			{Reviews of Modern Physics}\ }\textbf {\bibinfo {volume} {79}},\ \bibinfo
		{pages} {135} (\bibinfo {year} {2007})}\BibitemShut {NoStop}%
	\bibitem [{\citenamefont {Hu}\ \emph {et~al.}(2007)\citenamefont {Hu},
		\citenamefont {Peng}, \citenamefont {Li},\ and\ \citenamefont
		{Guo}}]{Hu2007OnLasers}%
	\BibitemOpen
	\bibfield  {author} {\bibinfo {author} {\bibfnamefont {Y.}~\bibnamefont
			{Hu}}, \bibinfo {author} {\bibfnamefont {X.}~\bibnamefont {Peng}}, \bibinfo
		{author} {\bibfnamefont {T.}~\bibnamefont {Li}}, \ and\ \bibinfo {author}
		{\bibfnamefont {H.}~\bibnamefont {Guo}},\ }\href {\doibase
		10.1016/j.physleta.2007.03.004} {\bibfield  {journal} {\bibinfo  {journal}
			{Physics Letters, Section A: General, Atomic and Solid State Physics}\
		}\textbf {\bibinfo {volume} {367}},\ \bibinfo {pages} {173} (\bibinfo {year}
		{2007})}\BibitemShut {NoStop}%
	\bibitem [{\citenamefont {Jalas}\ \emph {et~al.}(2013)\citenamefont {Jalas},
		\citenamefont {Petrov}, \citenamefont {Eich}, \citenamefont {Freude},
		\citenamefont {Fan}, \citenamefont {Yu}, \citenamefont {Baets}, \citenamefont
		{Popovi{\'{c}}}, \citenamefont {Melloni}, \citenamefont {Joannopoulos},
		\citenamefont {Vanwolleghem}, \citenamefont {Doerr},\ and\ \citenamefont
		{Renner}}]{Jalas2013WhatIsolator}%
	\BibitemOpen
	\bibfield  {author} {\bibinfo {author} {\bibfnamefont {D.}~\bibnamefont
			{Jalas}}, \bibinfo {author} {\bibfnamefont {A.}~\bibnamefont {Petrov}},
		\bibinfo {author} {\bibfnamefont {M.}~\bibnamefont {Eich}}, \bibinfo {author}
		{\bibfnamefont {W.}~\bibnamefont {Freude}}, \bibinfo {author} {\bibfnamefont
			{S.}~\bibnamefont {Fan}}, \bibinfo {author} {\bibfnamefont {Z.}~\bibnamefont
			{Yu}}, \bibinfo {author} {\bibfnamefont {R.}~\bibnamefont {Baets}}, \bibinfo
		{author} {\bibfnamefont {M.}~\bibnamefont {Popovi{\'{c}}}}, \bibinfo {author}
		{\bibfnamefont {A.}~\bibnamefont {Melloni}}, \bibinfo {author} {\bibfnamefont
			{J.~D.}\ \bibnamefont {Joannopoulos}}, \bibinfo {author} {\bibfnamefont
			{M.}~\bibnamefont {Vanwolleghem}}, \bibinfo {author} {\bibfnamefont {C.~R.}\
			\bibnamefont {Doerr}}, \ and\ \bibinfo {author} {\bibfnamefont
			{H.}~\bibnamefont {Renner}},\ }\href {\doibase 10.1038/nphoton.2013.185}
	{\bibfield  {journal} {\bibinfo  {journal} {Nature Photonics}\ }\textbf
		{\bibinfo {volume} {7}},\ \bibinfo {pages} {579} (\bibinfo {year}
		{2013})}\BibitemShut {NoStop}%
	\bibitem [{\citenamefont {Fujita}\ \emph {et~al.}(2000)\citenamefont {Fujita},
		\citenamefont {Levy}, \citenamefont {Osgood}, \citenamefont {Wilkens},\ and\
		\citenamefont {D{\"{o}}tsch}}]{Fujita2000WaveguideInterferometer}%
	\BibitemOpen
	\bibfield  {author} {\bibinfo {author} {\bibfnamefont {J.}~\bibnamefont
			{Fujita}}, \bibinfo {author} {\bibfnamefont {M.}~\bibnamefont {Levy}},
		\bibinfo {author} {\bibfnamefont {R.~M.}\ \bibnamefont {Osgood}}, \bibinfo
		{author} {\bibfnamefont {L.}~\bibnamefont {Wilkens}}, \ and\ \bibinfo
		{author} {\bibfnamefont {H.}~\bibnamefont {D{\"{o}}tsch}},\ }\href {\doibase
		10.1063/1.126284} {\bibfield  {journal} {\bibinfo  {journal} {Applied Physics
				Letters}\ }\textbf {\bibinfo {volume} {76}},\ \bibinfo {pages} {2158}
		(\bibinfo {year} {2000})}\BibitemShut {NoStop}%
	\bibitem [{\citenamefont {Gisin}\ \emph {et~al.}(2002)\citenamefont {Gisin},
		\citenamefont {Ribordy}, \citenamefont {Tittel},\ and\ \citenamefont
		{Zbinden}}]{Gisin2002QuantumCryptography}%
	\BibitemOpen
	\bibfield  {author} {\bibinfo {author} {\bibfnamefont {N.}~\bibnamefont
			{Gisin}}, \bibinfo {author} {\bibfnamefont {G.}~\bibnamefont {Ribordy}},
		\bibinfo {author} {\bibfnamefont {W.}~\bibnamefont {Tittel}}, \ and\ \bibinfo
		{author} {\bibfnamefont {H.}~\bibnamefont {Zbinden}},\ }\href {\doibase
		10.1103/RevModPhys.74.145} {\bibfield  {journal} {\bibinfo  {journal}
			{Reviews of Modern Physics}\ }\textbf {\bibinfo {volume} {74}},\ \bibinfo
		{pages} {145} (\bibinfo {year} {2002})}\BibitemShut {NoStop}%
\end{thebibliography}

%%%%%%%%%%%%%%


\vfill
\end{document}